\documentclass[11pt]{article}
\usepackage{latexsym}
\usepackage{amssymb,color}
\usepackage{amscd}
\usepackage{bbm}
\usepackage{fancybox}
\usepackage{cite}
\usepackage{amsmath,amsfonts,amsbsy}
\usepackage{graphicx}
\usepackage{epsfig}
\usepackage{psfrag}
\usepackage{comment}
\usepackage{tikz,tikz-cd}
\numberwithin{equation}{section}
\usepackage[font=footnotesize,labelsep=newline,labelfont=sc,justification=centering,
position=top]{caption}
\newcommand{\hoch}[1]{$\, ^{#1}$}

\usepackage{float}

\usepackage{multirow}                     
\usepackage{float}                          
\usepackage{lscape}                         
\usepackage{bm}

\definecolor{MyDarkBlue}{rgb}{0.15,0.15,0.45}
\definecolor{MyGreen}{rgb}{0.15,0.45,0.45}
\definecolor{MyPurple}{rgb}{0.55,0.25,0.55}
\usepackage[linktocpage=true]{hyperref}
\hypersetup{colorlinks=true,citecolor=MyPurple,linkcolor=MyPurple,urlcolor=MyGreen}

\usepackage[utf8]{inputenc}
\usepackage[T1]{fontenc}
\usepackage{lmodern}

\begin{document}

\begin{flushright}
\hfill{ \
\ \ \ \   ICCUB-15-008,  UG-15-67 \ \ \ \ }
\end{flushright}
\vskip 1.2cm

\begin{center}
{\Large \bf The Symmetries of the Carroll  Superparticle  }
\end{center}
\vspace{25pt}
\begin{center}
{\Large {\bf }}

\vspace{10pt}

{\Large Eric Bergshoeff\hoch{1},
Joaquim Gomis\hoch{2}
and Lorena Parra\hoch{1,3}
}

\vspace{10pt}

\hoch{1} {\it Van Swinderen Institute for Particle Physics and Gravity,
University of Groningen,\\
Nijenborgh 4, 9747 AG Groningen, The Netherlands}\\

\hoch{2} {\it Departament d'Estructura i Constituents de la Mat\`eria
and Institut de Ci\`encies del\\
Cosmos, Universitat de Barcelona, Diagonal 645, 08028 Barcelona, Spain}

\hoch{3} {\it Instituto de Ciencias Nucleares, Universidad Nacional
Aut\'onoma de M\'exico,\\
Apartado Postal 70-543, 04510 M\'exico, D.F., M\'exico}\\

\vspace{30pt}

\underline{ABSTRACT}

\end{center}

Motivated by recent  applications of Carroll symmetries we investigate the geometry of flat and curved (AdS)  Carroll space and the symmetries of a particle moving in such a space both in the bosonic as well as in the supersymmetric case. In the bosonic case we find that the Carroll particle possesses an infinite-dimensional symmetry which only in the flat case includes dilatations. The duality between the Bargmann and Carroll algebra, relevant for the flat case,  does not extend to the curved case.

In the supersymmetric case we study the dynamics of the ${\cal N}=1$ AdS Carroll superparticle. Only in  the flat limit we find that the action is invariant under an infinite-dimensional symmetry that includes a supersymmetric extension of the Lifshitz Carroll algebra with dynamical exponent $z=0$. We also discuss 
in the flat case the extension to ${\cal N}=2$ supersymmetry and show that
the flat ${\cal N}=2$ superparticle  is equivalent to the (non-moving) ${\cal N}=1$ superparticle and that therefore it is not BPS unlike its Galilei counterpart. This is due to the fact that in this case kappa-symmetry eliminates the linearized supersymmetry.

In an appendix we discuss the ${\mathcal N=2}$ curved case  in three dimensions only and show that there are two ${\cal N}=2$ theories that are physically different.

\vspace{15pt}

\thispagestyle{empty}

\vspace{15pt}

\vfill

\thispagestyle{empty}
\voffset=-40pt

\newpage

\tableofcontents


\newpage


\section{Introduction}

Space-time symmetries have played a central role in the understanding of various physical theories such as Newtonian Gravity, Maxwell's Electromagnetism, Special Relativity, General Relativity, Strings and Supergravity. Most of these models are based on relativistic symmetries.  An example of a model with non-relativistic symmetries is Newtonian Gravity which is based on the Galilei symmetries. Such non-relativistic symmetries arise  when the velocity of light is sent to infinity.

A formulation of non-relativistic gravity that is invariant under diffeomorphisms was introduced by Cartan \cite{Cartan:1923zea}, see also \cite{Havas:1964zza,anderson1967principles,Trautman,Kunzle1972,PhysRevD.22.1285}. This so-called Newton-Cartan gravity can be reformulated as a gauge theory of the Bargmann algebra \cite{DePietri:1994je,Andringa:2010it}. The interest in  Galilean-invariant theories with diffeomorphism invariance has increased recently due to their relation with  condensed matter systems \cite{Son:2005rv,Son:2013rqa,Geracie:2014nka},  see  also \cite{Jensen:2014wha,Banerjee:2015tga} and references therein. Galilean-invariant theories have also appeared recently in studies of Lifshitz holography \cite{Christensen:2013rfa,Hartong:2015wxa}.

Other non-relativistic theories such as non-relativistic superstrings and superbranes have been studied as special points  in the parameter space of M-theory\cite{Gomis:2000bd,Danielsson:2000gi}. Non-relativistic strings  have also attracted attention due to the fact that they appear as a possible soluble sector within string theory or M-theory \cite{Gomis:2004pw,Gomis:2005pg}.

A less well known example of a non-relativistic symmetry are the Carroll symmetries which arise when the velocity of light is sent to zero \cite{Levy1965}. In this sense the Carroll symmetries are the opposite to the Galilei symmetries.  This can also be seen by looking at the light cone which in the Carroll case, at each point of spacetime,  collapses to the time axis whereas  in the Galilei case  it coincides with the space axis. These Carroll symmetries have played an important role in recent investigations. For instance, theories with Carroll symmetries occur in  studies of  tachyon condensation \cite{Gibbons:2002tv}. More recently, they also have appeared in the study of warped conformal field theories \cite{Hofman:2014loa}.

A systematic investigation of the possible relativity groups\footnote{A relativity group is an invariance group of a physical theory that contains the generators of special relativity: time translations, spatial translations, boosts and spatial rotations.} was initiated  by Bacry and L\'evy-Leblond  \cite{Bacry:1968zf}. They showed that all these groups can be obtained by a contraction of the anti-de Sitter (AdS) and de Sitter (dS) groups\footnote{In this paper we will only consider the AdS case.}. As Table 1 shows there are three different types of contractions: the non-relativistic limit $c\rightarrow\infty$ of the AdS group leads to the Newton-Hooke (NH) group. The flat limit $R\rightarrow\infty$ leads to the Poincar\'e (P) group and the ultra-relativistic limit $c\rightarrow0$ leads to the AdS-Carroll (AC) group \cite{Levy1965}\footnote{Bacry and L\'evy-Leblond \cite{Bacry:1968zf} call this algebra the  para-Poincar\'e algebra.}.  In a second stage, the flat limit of the AdS-Carroll group and the ultra-relativistic limit of the Poincar\'e group leads to the Carroll (C) group while the non-relativistic limit of the Poincar\'e group and the flat limit of the Newton-Hooke group leads to the Galilean (G) group.

\begin{table}[ht]
\begin{center}
\begin{tikzcd}
   & &\arrow{ddll}[swap]{c\rightarrow0} AdS \arrow{dd}{R\rightarrow\infty}
   \arrow{ddrr}{c\rightarrow\infty}& &\\
   & & & &\\
  AC\arrow{ddr}[swap]{R\rightarrow\infty} & & \arrow{ddl}[swap]{c\rightarrow0} P
  \arrow{ddr}{c\rightarrow\infty} & &\arrow{ddl}{R\rightarrow\infty} NH \\
  & & & &\\
  & C & & G &
\end{tikzcd}
\end{center}
\caption{\sl The figure displays the different contractions of the AdS group. The different abbreviations are explained in the text.
 \label{table:1}}
\end{table}

All the algebras corresponding to the groups given in Table 1 contain  the same commutators  involving spatial rotations. These commutators are given by
\begin{eqnarray}
    \left[M_{ab}, M_{cd}\right]&=&2\eta_{a[c}M_{d]b}-2\eta_{b[c}M_{d]a}\,,\\ [.2truecm]
    [M_{ab},P_c]&=&2\delta_{c[b}P_{a]}\,, \hskip 1truecm
    [M_{ab},K_c] = 2\delta_{c[b}K_{a]}\,,
\end{eqnarray}
where  $a=1,\ldots,D-1$, for a $D$-dimensional space-time. The Galilean algebra can extended with a  central charge generator $Z$ to the so-called Bargmann algebra \cite{Bargmann:1954gh}. It has been recently shown that there is a  duality between this Bargmann algebra and the Carroll algebra by the exchange of $Z$ and the generator of time translations $H$ \cite{Duval:2014uoa}. Note that this duality does not extend to a duality between the Newton-Hooke and AdS-Carroll algebras. This is due to the expression for the commutator $[P_a,P_b]$, see Table 2.

The aim of this paper is to  study the general structure of the Carroll symmetries along the same lines as this has been done for the Galilean symmetries. This will be done in two stages. As a first step we will study the geometry of the empty Carroll space considering the coset $ G/H={\textrm {AC}} /{\textrm {Hom AC}},$ where {\textrm {Hom AC}} is the homogeneous part of the AC algebra. In a second step we will put a particle in this Carroll space and construct an action describing its dynamics.

More specifically, in   the first part of this paper we  consider the bosonic AC algebra. In particular,  we will construct the action of a particle invariant under the symmetries corresponding to this algebra using the method of non-linear realizations \cite{Coleman:1969sm,Callan:1969sn}. This so-called  AC particle reduces, in the limit that the AdS radius goes to infinity, to  the Carroll particle that we studied in our previous paper \cite{Bergshoeff:2014jla}. A characteristic feature of the free Carroll particle  is that it does not move \cite{old,Duval:2014uoa,Bergshoeff:2014jla}\footnote{If we consider two particles or a particle interacting with Carroll gauge fields the dynamics is non-trivial. The same phenomenon occurs in tachyon condensation when the tachyon interacts with a gauge field \cite{Gibbons:2002tv}.}. As we will see the AC particle does not move, but unlike the Carroll particle the momenta are not a constant of motion as a consequence of the AdS-Carroll symmetry. Another difference with the Carroll particle is that the mass-shell constraint depends on the coordinates of the AC space, therefore the AC particle `sees' the geometry. This is different from the Carroll case where the energy of the particle is equal to plus or minus the mass \cite{Duval:2014uoa,Bergshoeff:2014jla}. We find that only in the massless limit the mass-shell constraint coincides with the flat Carroll case. Using the AC particle action we will construct the Killing equations for the AC space. We find that the solution of the Killing equations produces an infinite-dimensional algebra that contains the symmetries of the AC algebra. The Lifshitz dilatations are not included in these symmetries. Only in the flat case the dilatations with $z=0$ are part of the infinite dimensional algebra.

\begin{table}[ht]
\begin{center}
{\small
\begin{tabular}{|c|c|c|c|c|c|}
\hline \rule[-1mm]{0mm}{6mm}
  &$[P_a,K_b]$ & $[H,K_a]$ & $[H,P_a]$   & $[P_a,P_b]$ & $[K_a,K_b]$ \\[.1truecm]
\hline \rule[-1mm]{0mm}{6mm}
AdS & $\delta_{ab}H$ & $P_a$ & $-\frac 1{R^2}K_a$ & $\frac 1{R^2}M_{ab}$ & $M_{ab}$\\[.1truecm]
\hline  \rule[-1mm]{0mm}{6mm}
Poincar\'e & $\delta_{ab}H$ & $P_a$ & 0 & 0 & $M_{ab}$ \\[.1truecm]
Newton-Hooke& $\delta_{ab} Z$  & $P_a$ & $-\frac 1{R^2}K_a$&$0$& 0 \\[.1truecm]
AdS-Carroll& $\delta_{ab}H$ & 0 & $-\frac 1{R^2}K_a$ & $\frac 1{R^2}M_{ab}$& 0 \\[.1truecm]
\hline  \rule[-1mm]{0mm}{6mm}
Galilei& $\delta_{ab} Z$ & $P_a$ & 0 & 0 & 0 \\[.1truecm]
Carroll&  $\delta_{ab}H$ & 0 & 0 & 0& 0 \\[.1truecm]
\hline
\end{tabular}
}
\end{center}
 \caption{\sl This table gives an overview of the algebras of the relativity groups that we consider in this paper. }\label{table:2}
\end{table}

In the second part of this paper we consider the supersymmetric extension of the Carroll algebras\,{\footnote{A first attempt in this direction was done in the unpublished notes \cite{old}.}}.
We first construct the $\mathcal N=1$ AC superalgebra in any dimension (see Tables \ref{table:2} and \ref{table:3}, where $Q$ stands for the generator of supersymmetry). A difference with respect to the supersymmetric Newton-Hooke  case is that we have a conventional supersymmetry algebra, where the energy and boost generators appear in the anti-commutator of the supersymmetries. The AC superalgebra in the flat limit contains the supersymmetric  extension of the `Lifshitz boost extended  Carroll algebra' introduced in appendix  B of \cite{Gibbons:2009me}. We construct the AC superparticle action both as the non-relativistic limit of the relativistic massive superparticle \cite{Casalbuoni:1976tz, Brink:1981nb} as well as  by applying the non-linear realization technique. As we will see the  ${\cal N}=1$ AC superparticle  like in the Relativistic and Galilean  case is non-BPS, i.e.~the  supersymmetries are non-linearly realized. We will study the super-Killing equations and we find in general an infinite-dimensional algebra of symmetries thereby extending the finite $\mathcal N=1$ super AC transformations.

\begin{table}[ht]
\begin{center}
{\small
\begin{tabular}{|c|c|c|c|}
\hline \rule[-1mm]{0mm}{6mm}
$\mathcal N=1$ & $[M_{ab},Q]$  &$[P_a,Q]$ &$\{Q_\alpha,Q_\beta\}$  \\[.1truecm]
\hline \rule[-1mm]{0mm}{6mm}
Newton-Hooke & $-\frac12\gamma_{ab} Q$ & $0$ &$2\delta_{\alpha\beta}Z$   \\[.1truecm]
AdS-Carroll & $-\frac12\gamma_{ab} Q$ & $\frac1{2R}\gamma_aQ$&$[\gamma^0C^{-1}]_{\alpha\beta}H+\frac 2R[\gamma^{a0}C^{-1}]_{\alpha\beta}K_a$   \\[.1truecm]
\hline
\end{tabular}
}
\end{center}
 \caption{\sl In this table we give the (anti-)commutators of the $\mathcal N=1$ Newton-Hooke and AdS-Carroll superalgebras that involve the generators $Q$ of supersymmetry. Note that  here is no duality between the two algebras. }\label{table:3}
\end{table}

\begin{table}[ht]
\begin{center}
{\small
\begin{tabular}{|c|c|c|c|c|c|}
\hline \rule[-1mm]{0mm}{6mm}
$\mathcal N=2$ & $[M_{ab},Q^\pm]$ &$[K_a,Q^+]$ &$\{Q_\alpha^+,Q_\beta^+\}$ & $\{Q_\alpha^+,Q_\beta^-\}$
   & $\{Q_\alpha^-,Q_\beta^-\}$   \\[.1truecm]
\hline \rule[-1mm]{0mm}{6mm}
Galilei& $-\frac12\gamma_{ab} Q^\pm$&  $-\frac12\gamma_{a0}Q^-$ &$[\gamma^0C^{-1}]_{\alpha\beta}H$
    & $[\gamma^aC^{-1}]_{\alpha\beta}P_a$ & $2[\gamma^0C^{-1}]_{\alpha\beta}Z$ \\[.1truecm]
    Carroll& $-\frac12\gamma_{ab} Q^\pm$ & 0 &$\frac12[\gamma^0C^{-1}]_{\alpha\beta}(H+2Z)$ & 0
    & $\frac12[\gamma^0C^{-1}]_{\alpha\beta}(H-2Z)$  \\[.1truecm]
\hline
\end{tabular}
}
\end{center}
 \caption{\sl In this table we give the (anti-)commutators of the $\mathcal N=2$  Galilei and Carroll supersymmetry algebras. Note that there is no duality between these two algebras. }\label{table:4}
\end{table}

Inspired by the relativistic and Galilei case we will investigate whether the ${\cal N}=2$ Carroll superparticle is BPS or not. For simplicity we restrict to the three-dimensional case. We first construct the  $\mathcal N=2$ Carroll superalgebra as a contraction of the $\mathcal N=2$ Poincar\'e superalgebra. This leads to the result given in Table \ref{table:4}. We see that, unlike in the bosonic case, there is no duality in the supersymmetric case. Next, we construct the  action for the ${\cal N}=2$ Carroll superparticle. This action has two terms, one of them is a Wess-Zumino (WZ)  term. If we properly choose the coefficients of the two terms we find a so-called kappa gauge symmetry \cite{deAzcarraga:1982dw,Siegel:1983hh} that kills half of the fermions. This gauge symmetry has the form of a St\"uckelberg symmetry, similar to what we found in the Galilean case \cite{Gomis:2004pw,Bergshoeff:2014gja}. We find that after fixing the kappa-symmetry the super-Carroll action reduces to the action we found in the $\mathcal N=1$ case.  The linearly realized supersymmetry acts trivially on all the fields and therefore the ${\cal N}=2$ Carroll superparticle reduces to the ${\cal N}=1$ Carroll superparticle and hence is not BPS. This is rather different from the ${\cal N}=2$ super-Galilei case were BPS particles do exist. The main difference between the super-Carroll and super-Galilei cases comes from the kappa symmetry transformations, in the former case it eliminates the linearized supersymmetry and it the last case it does not.

In a separate Appendix we extend our investigations to the ${\cal N}=2$ curved  case and consider the Carroll contraction of the so-called (p,q) AdS superalgebras \cite{Achucarro:1987vz} for  the particular cases of $(p,q)=(2,0)$ and $(p,q)=(1,1)$. The (2,0) and (1,1)  AdS Carroll algebras are not isomorphic. We find that the associated particle actions are rather different. While  in the (2,0) case we have kappa-symmetry, we find that  this is not the case in the (1,1) case. The two models have different degrees of freedom.

This paper is organized as follows. In section 2 we discuss the bosonic free AC particle thereby extending our previous analysis \cite{Bergshoeff:2014jla} to the curved case. In particular, we construct the action and investigate the Killing equations. In section 3 we consider the  $\mathcal N=1$ AC superparticle. At the end of this section we discuss the flat limit. Finally, in section 4 we investigate the $\mathcal N=2$ Super Carroll particle. Our conclusions are presented in section 5. Some technical details and the extension of the $\mathcal N=2$ Super Carroll particle to the curved case, for three dimensions only,  are given in three Appendixes.

\section{The Free AdS Carroll Particle}

Before discussing the supersymmetric case we will first study in this section different aspects of the free AdS Carroll (AC)  particle.

\subsection{The AdS Carroll Algebra}
In order to write the commutators corresponding to the AC algebra, we will start with the contraction of the $D$-dimensional AdS algebra. The basic commutators are given by $(A=0,1,\ldots,D-1)$
\begin{align}
    \left[M_{AB}, M_{CD}\right]&=2\eta_{A[C}M_{D]B}-2\eta_{B[C}M_{D]A}\,,\\
    [M_{AB},P_C]&=2\eta_{C[B}P_{A]}\,,\hspace{1cm}
    [P_A,P_B]=\frac{1}{R^2}M_{AB}\,,
\end{align}
where  $R$ is the AdS radius. Here $P_A$ and $M_{AB}$ are the (anti-hermitian) generators of space-time translations and Lorentz rotations, respectively.

To make the Carroll contraction we rescale the generators  with a parameter $\omega$ as follows  \cite{Levy1965,Bacry:1968zf}:
\begin{equation}
\begin{aligned}
    P_0&=\frac\omega2 H\,,\hspace{1cm}
    M_{a0}=\omega K_a\,.\\
\end{aligned}
\end{equation}
Taking the limit $\omega\rightarrow\infty$ we find that the commutators corresponding to the $D$-dimensional AC algebra are given by ($a=1,\ldots,D-1$):
\begin{align}
    \label{AdSCarrAlgBos}
    \left[M_{ab}, M_{cd}\right]&=2\eta_{a[c}M_{d]b}-2\eta_{b[c}M_{d]a}\,,\hspace{1cm}
    [M_{ab},K_c]=2\delta_{c[b}K_{a]}\,, \\
    \left[M_{ab},P_c\right]&=2\delta_{c[b}P_{a]}\,, \hspace{3.6cm}
    [P_a,K_b]=\frac12\delta_{ab}H\,,\\
    [P_a, P_b]&=\frac1{R^2}M_{ab}\,,\hspace{3.7cm}
    [P_a,H]=\frac 2{R^2}K_a\,.
\end{align}
Notice that the commutation relations of space-time translation coincide with the same commutation relations of the AdS algebra. The difference between the AdS and AC algebra is in the different commutation relations that involve the boost generators. Note that this is not the case for the Newton-Hook algebras.

The AC algebra can be expressed in terms of the left invariant Maurer-Cartan 1-forms  $L^a$, which satisfy the Maurer-Cartan equations $dL^C-\frac12\,{f^C}_{AB}L^BL^A=0$. Explicitly, these equations read
\begin{align}
    dL_H +\frac12 L_P^{\;a}\,L_K^{\;a}&=0 \,,\hspace{3.2cm}
    d L^{\;a}_P- 2L_P^{\;b}\,L_M^{\;ab}=0\,,\\
    d L^{\;a}_K- 2L_K^{\;b}\,L_M^{\;ab}&=\frac2{R^2}L_H{L_P}^a \,,\hspace{1.45cm}
    d L_M^{\;ab}- 2L_M^{\;ca}\,L_M^{\;cb}=\frac1{2R^2}{L_P}^b{L_P}^a\,.
\end{align}

\subsection{Non-Linear Realizations}

In  this subsection  we apply the method of non-linear realizations \cite{Coleman:1969sm,Callan:1969sn} and  use the algebra \eqref{AdSCarrAlgBos} to construct the action of the AC  particle.

We consider the coset $ G/H={\textrm {AC}} /{\textrm {SO(D-1)}}$ and the coset element $g= g_0\,U, \label{gold20}$ where $g_0=e^{Ht} e^{ P_a x^a}$ is the coset representing  the AC space and $U=e^{ K_a v^a}$ is a general Carroll boost. The $x^{a}\, (a=1,...D-1)$ are the Goldstone bosons of broken translations, $t$ is the Goldstone boson of the unbroken time translation\footnote{The unbroken translation $P_0$ generates via a right action  \cite{McArthur:1999dy}\cite{Gomis:2006xw} a transformation which is equivalent to the world-line diffeomorphisms.} and  $U$ is parametrized by the Goldstone bosons of the broken Carroll boost transformations.

The reason to consider the coset element in terms of $g_0$ and $U$ is because in this way we have that for a general symmetric space-time $g_0$ is the coset element representing  the `empty' space-time, while $U$ represents the broken symmetries that are due to the presence of a dynamical object, in our case a particle, in the `empty' space-time. For the case of a particle $U$ is given by the general rotation that mixes the `longitudinal' time direction with the `transverse' space directions, i.e.~the Carroll boosts. If we would like to consider as a dynamical object a p-brane, we should consider as $U$ the general rotations that mix the longitudinal and tranverse directions  \cite{Gomis:2006xw}.

Returning to the AC particle, it is interesting to write out the Maurer-Cartan form $\Omega_0$ associated to the AC space
\begin{equation}
\Omega_0= g_0^{-1}dg_0=
H e^0+ P_a e^a  + K_a \omega^{a0}+
M_{ab}\omega^{ab}\,,
\end{equation}
where $(e^0,e^a)$ and $(\omega^{a0},\omega^{ab})$ are the space and time components of the Vielbein and  spin connection 1-forms of the AdS space, respectively. If we parametrize the AdS space as $e^{Ht} e^{ P_a x^a}$, the Vielbein and spin-connection 1-forms corresponding to the AC space are given by
\begin{equation}
\begin{aligned}
\label{3DLsAdSBos1}
    e^0& = d t \cosh\frac xR
          \,,\\
    e^a &=\frac Rxdx^a\sinh\frac xR
              +\frac 1{x^2}x^ax^bdx_b\Big(1-\frac Rx \sinh \frac xR\Big) \,,\\
    \omega^{a0} &= -\frac 2{xR}dtx^a\sinh\frac xR  \,,\\
    \omega^{ab}&=\frac1{2 x^2}(x^b dx^a- x^adx^b)\Big(\cosh \frac xR-1 \Big)\,.\\
    \end{aligned}
\end{equation}
These 1-forms satisfy the structure equations
\begin{align}
\label{eq1Carroll}
    de^0 +\frac12 e^{\;a}\,\omega^{\;a0}&=0 \,,\hspace{3.2cm}
    d e^{\;a}- 2e^{\;b}\,\omega^{\;ab}=0\,,\\
\label{eq2Carroll}
    d e^{\;a}- 2\omega^{\;b0}\,\omega^{\;ab}&=\frac2{R^2}e^0{e}^a \,,\hspace{1.75cm}
    d \omega^{\;ab}- 2\omega^{\;ca}\,\omega^{\;cb}=\frac1{2R^2}{e}^b{e}^a\,.
\end{align}
We see that the Vielbein satisfies  the torsionless condition and  that the AC space, like the ancestor AdS space,  has constant negative curvature.

We now insert a particle in the empty AC space and consider the Maurer-Cartan form of the combined system:
\begin{equation}
\Omega=g^{-1}dg=U^{-1}\Omega_0 U+ U^{-1}dU\,.
\end{equation}
In order to derive an expression for $\Omega$ we  need to know how the space-time translation generators and the  boost generators transform under a general Carroll boost:
\begin{equation}
\begin{aligned}
\label{Utrans}
    U^{-1}\,H\, U& = H+\frac 12 v^a  P_a
    \,,\\
    U^{-1}\,P_a\, U &=P_a\,,\\
     U^{-1}\,K_a\, U &=K_a\,,\\
     U^{-1}\,M_{ab}\, U &=M_{ab}+v_bK_a-v_aK_b\,.
    \end{aligned}
\end{equation}
We have also $U^{-1}dU=dv^aK_a$. Using these formulae we find that the Maurer-Cartan form $\Omega$ is given by
\begin{equation}
\begin{aligned}
\label{3DLsAdSBos2}
    L_H& = e^0+\frac 12 v_a  e^a\,,\\
    {L_P}^a &=e^a \,,\\
    {L_K}^a &= \omega^{0a}+ dv^a
               +2v_b\,\omega^{ab}\,,\\
      {L_M}^{ab}&= \omega^{ab}\,.
\end{aligned}
\end{equation}
We note that that the Maurer-Cartan forms of space-time translations can be written in matrix-form as follows:
\begin{equation}
  \big ( \begin{array}{cc} L_H, & {L_P}^a \end{array} \big)=  \big ( \begin{array}{cc} e^0, & e^a \end{array} \big) \begin{pmatrix} 1 & 0 \\ \frac12 v_a & 1 \end{pmatrix}  \,.
\end{equation}
The matrix appearing at the right-hand-side  is the most general Carroll boost in the vector representation.

We now proceed with the construction of an action of the AC particle. An action  with the lowest number of derivatives is obtained  by taking the pull-back of all the $L$'s that are invariant under rotations, see for example \cite{Gomis:2006xw}. In this way we  obtain the following action:
\begin{equation}
\label{AdSAction3DBos}
\begin{aligned}
    S&=M\int(L_H)^* =M\int \Big(e^0+\frac 12 v_a  e^a\Big)^*\\
     &=M\int d\tau \left( \dot t \cosh\frac xR+\frac R{2x}v_a\dot x^a\sinh\frac xR
         +\frac1{2x^2}x^bv_b x_a \dot x^a \Big(1-\frac Rx\sinh\frac xR\Big) \right) \,.
\end{aligned}
\end{equation}
This action is invariant under the following transformation rules with constant parameters $(\zeta,    a^i, \lambda^i, \lambda^i_j)$ corresponding to time translations, spatial translations, boosts and spatial rotations, respectively:
\begin{equation}
\begin{aligned}
\label{AdSBosTransf}
    \delta t&=-\zeta+\frac R{2x} \lambda^k x_k \tanh \frac xR
              +\frac t{Rx}a^k x_k\tanh\frac xR\,,\\[.1truecm]
    \delta x^i&= -\frac 1{x^2}\left(x^ia^k x_k
                 -\frac xR \coth\frac xR (x^ia^kx_k-a^ix^2)\right)
                 -2\lambda^{i}_{\,k}\,x^k \,,\\[.1truecm]
    \delta v^i&= -\lambda^i-\frac1{x^2}\lambda^kx_kx^i\text{sech}\frac xR
     \left(1-\cosh \frac xR\right)
               -2\lambda^{i}_{\,j}\,v^j -\frac{2t}{R^2}a^i\\[.1truecm]
               &\hspace{.4cm}-\frac{2t}{R^2x^2}x^ia^kx_k\,
               \text{sech}\frac xR\left(1-\cosh\frac xR\right)
               +\frac2{Rx}v_ba^{[i}x^{b]}\text{csch}\frac xR
               \left( 1-\cosh\frac xR\right)\,.\\
\end{aligned}
\end{equation}
The equations of motion for $t$, $x^a$ and $v^a$ read
\begin{equation}
\begin{aligned}
    0&=\frac1{xR} x^a\dot x_a \sinh\frac xR\,,\\[.1truecm]
    0&=-\frac R{2x}\dot x^a\sinh\frac xR
       -\frac1{2x^2}x^ax_b\dot x^b\Big(1-\frac Rx\sinh\frac xR\Big)\,,\\[.1truecm]
    0&=\frac R{2x}\dot v_a\sinh\frac xR-\frac1{xR}\dot t x_a\sinh\frac xR
       +\frac1{2x^2}x_ax^b\dot v_b \Big(1-\frac Rx\sinh\frac xR\Big)\\[.1truecm]
    &\hspace{.5cm}+\frac {\dot x^b}{2x^2}(v_a x^b-x_av_b)\Big(\cosh\frac xR-1\Big)\,.
\end{aligned}
\end{equation}
These equations imply that
\begin{equation}
\begin{aligned}
\label{ecsmovAdSbos}
    &\dot x^a=0\,,\\
    & \frac1{xR}\dot t x_a\sinh\frac xR=\frac R{2x}\dot v_a\sinh\frac xR
           +\frac1{2x^2}x_ax^b\dot v_b \Big(1-\frac Rx\sinh\frac xR\Big)\,.
\end{aligned}
\end{equation}
Notice that the evolution of  $v^a$ is non-trivial. If we take the limit $R\rightarrow\infty$ we recover the flat bosonic equations of motion $\dot x_a=\dot v_a=0$ and therefore a trivial dynamics for both $x^a, v^a$ \cite{Bergshoeff:2014jla}.

The energy and spatial momenta of the free AC particle are given by
\begin{equation}
\begin{aligned}
\label{MomentaAdS}
    E&= -\frac{\partial\mathcal L}{\partial\dot t} = -M\cosh\frac xR \,,\\[.1truecm]
    p_a&=\frac{\partial\mathcal L}{\partial \dot x_a}
    =M\left[\frac R{2x} v_a \sinh\frac xR
       +\frac 1{2x^2}x_ax^bv_b\Big(1-\frac Rx\sinh\frac xR\Big) \right]\,.\\
\end{aligned}
\end{equation}
They satisfy the  constraint
\begin{equation}
E^2-M^2\cosh^2\frac xR=0\,.
\end{equation}

The canonical action of the AC  particle is given by\,\footnote{Alternatively, we can obtain this action by taking the Carroll limit of the canonical action of a massive particle in AdS, see appendix A.}

\begin{equation}
\label{AdSBosActNLR}
    S=\int d\tau\left[-E \dot t+ p_a \dot{x}^a
                            -\frac e2\left(E^2-M^2\cosh^2 \frac xR\right)\right]\,.
\end{equation}
Note that if we calculate $\dot p_a$ and impose both equations of motion \eqref{ecsmovAdSbos}
we obtain
\begin{equation}
\begin{aligned}
    \dot p_a=\frac M{Rx} \dot t x_a \sinh\frac xR
            =\frac{eM^2}{Rx}x_a\cosh\frac xR\sinh\frac xR\,.
\end{aligned}
\end{equation}
In the last step we have used  that $\dot t=-eE=eM\cosh\frac xR$, see eq.~\eqref{ecsmovAdSbosHam}. This is the same result one finds using the Hamiltonian form given in eq.~\eqref{ecsmovAdSLim}.

\subsection{The Killing Equations of the AdS Carroll Particle}

In order to find the Killing symmetries of the AC space, it is convenient to consider the symmetries of the canonical action \eqref{AdSBosActNLR}. The basic Poisson brackets of the canonical variables occurring in the action \eqref{AdSBosActNLR} are given by
\begin{equation}
    \{E,t\}=1\,,\qquad\{e,\pi_e\}=1\,,\qquad
    \{x_i,p_j\}=\delta_{ij}\,.
\end{equation}
This leads to the following equations of motion:
\begin{equation}
\begin{aligned}
\label{ecsmovAdSbosHam}
    &\dot t=-eE\,,\hspace{.5cm}
    \dot x^i=0\,,\hspace{.5cm}
    \dot E=0\,,\hspace{.5cm}
    \dot p^i=\frac{eM^2}{2Rx}x^i \sinh \frac {2x}{R}\,,\\
    &\hspace{1.5cm}\dot\pi_e=-\frac12\left(E^2-M^2\cosh^2\frac xR\right)\,,\hspace{.5cm}
    \dot e =\lambda\,.
\end{aligned}
\end{equation}
Here $\lambda=\lambda(\tau)$ is an arbitrary function and $\pi_e$ is constrained by $\dot\pi_e=0$.

We take as the generator of canonical transformations
\begin{equation}
    G=-E\xi^0(t,\vec x, e)+p_i\,\xi^i(t,\vec x, e)+\gamma(t,\vec x, e)\pi_e\,,
\end{equation}
where $\xi^0=\xi^0(t,\vec x, e),\xi^i=\xi^i(t,\vec x,e)$ and
$\gamma=\gamma(t,\vec x,e)$.
The condition that this generator generates a Noether symmetry is that it is a constant of motion and it leads to the following restrictions:
\begin{equation}
\begin{aligned}
   \dot G &=0 = -E(\dot t \partial_t\xi^0+\dot e\partial_e\xi^0)+\dot p_i \xi^i
               + p_i (\dot t \partial_t\xi^i+\dot e\partial_e\xi^i)+\gamma\dot\pi_e\\[.2truecm]
               &=eE^2\partial_t\xi^0-\lambda E \partial_e\xi^0
               + \frac{eM^2}{2Rx}x_i\xi^i\sinh\frac{2x}{R}\\[.2truecm]
               &\ \ \ -eEp_i\partial_t\xi^i+\lambda p_i\partial_e\xi^i
               -\frac\gamma 2\left(E^2-M^2\cosh^2\frac xR\right)\,.
\end{aligned}
\end{equation}
From this equation we deduce the following equations describing the symmetries of the AC space:
\begin{equation}
\begin{aligned}
    &\hspace{1.2cm}\partial_e \xi^0=0\,,\hspace{.5cm}
    \partial_e\xi^i=0\,,\hspace{.5cm}\partial_t\xi^i=0\,,\\
    &\gamma=2e\partial_t\xi^0\,,\hspace{1cm}
    \frac e{xR}x_i\xi^i\sinh\frac xR+\frac12\gamma\cosh\frac xR=0\,.
\end{aligned}
\end{equation}
The last two equations can be combined into the single condition
\begin{equation}
\begin{aligned}
    \partial_t\xi^0=-\frac 1{xR}x_i\xi^i\tanh\frac xR\,.
\end{aligned}
\end{equation}
The generator $G$ is given by
\begin{equation}
    G=-E\xi^0(t,\vec x)+p_i\,\xi^i(\vec x)+\gamma(t,\vec x, e)\pi_e\,.
\end{equation}

From the variation of the momenta we can obtain the transformation rules for $v_i$ as follows. First, we use that
\begin{eqnarray}
    &&\delta p_i=\{p_i,G\}
    =\{p_i,-E\xi^0(t,\vec x)+p_i\,\xi^i(\vec x)+2e\partial_t\xi^0(t,\vec x)\pi_e\}\nonumber\\[.1truecm]
    &&\ \ \ \ \ =E\partial_i\xi^0-p_k\partial_i\xi^k-2e\partial_t\partial_i\xi^0\pi_e\,.
\end{eqnarray}
Next, using eq.~\eqref{MomentaAdS} and $\pi_e=0$ we obtain
\begin{equation}
    \delta p_i=-M\cosh\frac xR\partial_i\xi^0-M\bigg[\frac R{2x}v_i\sinh\frac xR
                + \frac 1{2x^2}x_i x^bv_b
                   \Big(1-\frac Rx\sinh\frac xR\Big)\bigg]\partial_i\xi^k\,.
\end{equation}
Finally, using the expression for $p_i$ given in eq.~\eqref{MomentaAdS}, we obtain the following transformations of the variables $v_i$:
\begin{equation}
\begin{aligned}
\label{Transf_vAdS}
    \delta v_i=&-\frac{2x}{R}\partial_i\xi^0\coth\frac xR-v^a\partial_i\xi_a
            -\frac1{Rx}v_bx^bx_k\partial_i\xi^k\Big(1-\frac Rx\sinh\frac xR\Big)\\[.1truecm]
            &+\frac 2{Rx}\coth\frac xR\Big(1-\frac Rx\sinh\frac xR\Big)x_i
            x_a\partial^a\xi^0
             +\frac1{Rx}\Big(\frac Rx-\coth\frac xR\Big)v_ix_b\xi^b\\[.1truecm]
            &-\frac1{Rx}\Big(\frac Rx-\frac{R^2}{x^2}\sinh\frac xR\Big)
            \Big(x_ix_b\xi^b-\frac1{x^2}x_ix_a\partial^a\xi_kv^k\Big)\\[.1truecm]
            &+\frac1{Rx^3}\text{csch}\frac xR
            \Big(-\frac Rx\sinh\frac xR-\frac{R^2}{x^2}\sinh^2\frac xR+1+\cosh\frac
            xR\Big)
            x_ix_b\xi^bx_kv^k\\[.1truecm]
            &+\frac 1{Rx} \text{csch}
            \frac xR \Big(-2\frac Rx\sinh\frac xR-\frac{R^2}{x^2}\sinh^2\frac xR+1
                          +\cosh\frac xR\Big)
            x_ix_b\xi^bx_kv^k\\[.1truecm]
            &+\frac 1{Rx}\text{csch}\frac xR\Big(\frac Rx\sinh\frac xR-1\Big)
             \xi_ix_b v^b\,.
\end{aligned}
\end{equation}
We see that the free Carroll particle in an AdS background has an infinite-dimensional symmetry. A possible solution to these equations is given by eq.~\eqref{AdSBosTransf} which are the symmetry transformations of the Carroll group. We do not find any Lifshitz dilatations in this case i.e., a transformation with parameters  $\xi^i=x^i, \xi^0=zt$.

\subsubsection{The Massles Limit}

Using the canonical action
\begin{equation}
    S=\int d\tau\left[-E \dot t+ p_a \dot{x}^a
                            -\frac e2\left(E^2-M^2\cosh^2 \frac xR\right)\right]\,,
\end{equation}
it is straightforward to take the massless limit $M\rightarrow0$ and obtain the action
\begin{equation}
    S=\int d\tau\left (-E \dot t+ p_a \dot{x}^a -\frac e2E^2\right)\,.
\end{equation}
We see that in the massless limit the R-dependence of the AC particle has disappeared. This means that the {\sl massive} Carroll particles are affected by the geometry but the {\sl massless} Carroll particles are not. Consequently, in the massless limit there is no difference between particles in an AdS or flat background. Furthermore, the isometries should be given by the most general conformal Carroll group as it was analyzed in \cite{Bergshoeff:2014jla}. In this case dilatations are included i.e., with parameters $\xi^i=x^i$, $\xi^0=zt$.

\section{The \texorpdfstring{$\mathcal N=1 $}{N=1} AdS Carroll Superparticle}

In this section we extend our investigations to the $\mathcal N=1$ supersymmetric case and consider the AC superparticle.

\subsection{The \texorpdfstring{$\mathcal N=1$}{N=1} AdS Carroll Superalgebra}

We  start by taking the contraction of the $D$-dimensional $\mathcal N=1$ AdS algebra. The basic commutators are given by $(A=0,1,\dots,D-1)$
\begin{equation}
\begin{aligned}
    \left[M_{AB}, M_{CD}\right]&=2\eta_{A[C}M_{D]B}-2\eta_{B[C}M_{D]A}\,,\\
    [M_{AB},P_C]&=2\eta_{C[B}P_{A]}\,,\hspace{4.7cm}
    [P_A,P_B]=4x^2M_{AB}\,,\\
    [M_{AB},Q]&=-\frac12\gamma_{AB}Q\,,\hspace{4.95cm}
    [P_A,Q]=\frac{1}{2R}\gamma_A Q\,,\\
    \{Q_\alpha,Q_\beta\}&=2[\gamma^AC^{-1}]_{\alpha\beta}P_A
                              +\frac{1}{R}[\gamma^{AB}C^{-1}]_{\alpha\beta}M_{AB}\,,
\end{aligned}
\end{equation}
where $R$ is the AdS radius and $P_A,M_{AB}$ and $Q_\alpha$ are the generators of space-time translations, Lorentz rotations, and supersymmetry transformations, respectively. The bosonic generators $P_A$ and $M_{AB}$ are anti-hermitian while de fermionic generator $Q_\alpha$ is hermitian.

To make the Carroll contraction we rescale the generators with a parameter $\omega$ as follows:
\begin{equation}
\begin{aligned}
    P_0&=\frac\omega2 H\,,\hspace{1cm}
    M_{a0}=\omega K_a\,,\hspace{1cm}
    Q=\sqrt\omega\tilde Q\,;\hspace{1cm}
    a=1,2,\ldots,D-1
\end{aligned}
\end{equation}
Taking the limit $\omega\rightarrow\infty$ and dropping the tildes on the $Q$ we get the following $\mathcal N=1$ AdS Carroll superalgebra:
\begin{equation}
\begin{aligned}
    \label{AdSCarrAlgN1}
    \left[M_{ab},P_c\right]&=2\delta_{c[b}P_{a]}\,, \hspace{1cm}
    [M_{ab},K_c]=2\delta_{c[b}K_{a]}\,, \\
    [P_a, P_b]&=\frac1{R^2}M_{ab}\,,\hspace{1cm}
    [P_a,K_b]=\frac12\delta_{ab}H\,,\hspace{1cm}
    [P_a,H]=\frac 2{R^2}K_a\\
    [P_a,Q]&=\frac 1{2R}\gamma_a Q\,,\hspace{1cm}
    [M_{ab},Q]=-\frac 12 \gamma_{ab}Q\,,\\
    \{ Q_\alpha, Q_\beta\}&=[\gamma^0C^{-1}]_{\alpha\beta}H
                            +\frac 2R[\gamma^{a0}C^{-1}]_{\alpha\beta}K_a\,.
\end{aligned}
\end{equation}
The Maurer-Cartan equation $dL^C-\frac12\,{f^C}_{AB}L^BL^A=0$ in components reads
\begin{equation}
\begin{aligned}
    dL_H=& -\frac12 L_P^{\;a}\,L_K^{\;a}-\frac12 \bar L_Q\gamma^0L_Q \,,\hspace{3.3cm}
    d L^{\;a}_P= 2L_P^{\;b}\,L_M^{\;ab}\,,\\
    d L^{\;a}_K=& 2L_K^{\;b}\,L_M^{\;ab}+\frac2{R^2}L_H{L_P}^a
                  -\frac1R\bar L_Q\gamma^{a0}L_Q\,,\hspace{1.15cm}
    d L_M^{\;ab}=2L_M^{\;ca}\,L_M^{\;cb}+\frac1{2R^2}{L_P}^b{L_P}^a\,, \\
    d L_Q=&\frac12 \gamma_{ab} L_Q \,L_M^{\;ab}-\frac1{2R}\gamma_aL_Q{L_P}^a\,.
\end{aligned}
\end{equation}

\subsection{Superparticle Action}

We now use the algebra \eqref{AdSCarrAlgN1} to construct the action of the AC superparticle with the coset
\begin{equation}
 \frac G H=\frac {\mathcal N=1\textrm { AdS Carroll}}{\textrm {SO(D-1)}}
\end{equation}
that is locally parametrized as $g= g_0\,U,$ where $g_0=e^{Ht} e^{ P_a x^a} e^{Q_\alpha\theta^\alpha}$ is the coset representing  the `empty' curved AC Carroll superspace and $U=e^{ K_a v^a}$ is a general Carroll boost representing the particle inserted in the empty space. The Maurer-Cartan form $\Omega_0$ associated to the empty AC superspace is given by
\begin{equation}
\Omega_0= g_0^{-1}dg_0=
H E^0+ P_a E^a  + K_a \omega^{a0}+
M_{ab}\omega^{ab}-{\bar Q} E\,,
\end{equation}
where $(E^0,E^a,E_\alpha)$ and $(\omega^{a0},\omega^{ab})$ are the time and space components of the supervielbein and the spin connection of super-AdS if we parametrize the AdS superspace as  $e^{Ht} e^{ P_a x^a}e^{Q_\alpha\theta^\alpha}.$ The explicit expressions for these components are given by
\begin{equation}
\begin{aligned}
\label{3DEsAdSN1}
    E^0& = d t \cosh\frac xR-\frac12\bar{\theta}\gamma^0d\theta
     -\frac 1{2}\omega^{ab}\bar\theta\gamma_{ab}\gamma^0\theta\,,\\
    E^a &=\frac Rxdx^a\sinh\frac xR
              +\frac 1{x^2}x^ax^bdx_b\Big(1-\frac Rx \sinh \frac xR\Big) \,,\\
    \omega^{a0} &= -\frac 2{xR}dtx^a\sinh\frac xR
                   -\frac 1R\bar \theta \gamma^{a0}d\theta
                   -\frac1{2R^2}\bar\theta\gamma_{ab}\gamma^0\theta E^b\,,\\
    \omega^{ab}&=\frac1{2 x^2}(x^b dx^a- x^adx^b)\Big(\cosh \frac xR-1 \Big)\,,\\
    E_\alpha&= d\theta_\alpha -\frac 1{2R}[\gamma_a\theta]_\alpha
            E^a +\frac12\omega^{ab}[\gamma_{ab}\theta]_\alpha\,.
    \end{aligned}
\end{equation}
In this case we have torsion given by $T_0=-\frac12 \bar E^\alpha\gamma^0 E_\alpha$ and a non-vanishing spin connection. The Maurer-Cartan form for the $\mathcal N=1$ AC superparticle inserted in the AC superspace is given by
\begin{equation}
\Omega=g^{-1}dg=U^{-1}\Omega_0 U+ U^{-1}dU\,,
\end{equation}
where
\begin{equation}
\begin{aligned}
\label{3DLsAdSN1}
    L_H& = E^0+\frac 1{2}v_aE^a\,,\\
    {L_P}^a &=E^a \,,\\
    {L_K}^a &= \omega^{a0}+dv^a
               +2{v_b}\,\omega^{ab}\,,\\
    {L_M}^{ab}&=\omega^{ab}\,,\\
    {L_Q}_\alpha&=E_\alpha\,.\\
\end{aligned}
\end{equation}
Note that the Maurer-Cartan forms of the spacetime supertranslations can be written in matrix form in terms of the Supervielbein components of the AC superspace as follows:
\begin{equation}
\label{matrixsupertrans}
  \big ( \begin{array}{ccc} L_H, & {L_P}^a,&  {L_Q}_{\alpha}\end{array} \big)=  \big ( \begin{array}{ccc} E^0, & E^a, & E_\alpha\end{array} \big) \begin{pmatrix} 1 & 0 & 0\\ \frac12 v_a & 1& 0\\ 0 & 0 &1 \end{pmatrix}  \,.
\end{equation}
Like in the bosonic case the Maurer-Cartan forms of the supertranslations of the AC superparticle can be obtained from the Maurer-Cartan forms of the AC superspace by a matrix representation of the Carroll boost.

The action of the $\mathcal N=1$ AC superparticle is given by the pull-back of all the $L$'s that are invariant under rotations:
\begin{equation}
\label{AdSAction3DN1}
\begin{aligned}
    S&=M\int(L_H)^* =M\int \Big(E^0+\frac 12 v_a  E^a\Big)^*=\\
     &=M\int d\tau \left( \dot t \cosh\frac xR+\frac R{2x}v_a\dot x^a\sinh\frac xR
         +\frac1{2x^2}x^bv_b x_a \dot x^a \Big(1-\frac Rx\sinh\frac xR\Big)
         - \frac12 \bar\theta\gamma^0 \dot\theta \right.\\
         &\hspace{2.2cm} -\frac 1{4x^2}x^b\dot x^a \bar\theta\gamma_{ab}\gamma^0\theta
         \left.\Big(\cosh\frac xR -1\Big)\right)\,.\\
\end{aligned}
\end{equation}
The equations of motion corresponding to this action can be written as follows
\begin{equation}
\begin{aligned}
\label{AdSN1eqsmov}
    \dot x^i&=0\,,\hskip 2truecm
    \dot \theta = 0\,,\\[.1truecm]
    &\hspace{-.5cm} \frac1{xR}\dot t x_a\sinh\frac xR=\frac R{2x}\dot v_a\sinh\frac xR
               +\frac1{2x^2}x_ax^b\dot v_b\Big(1-\frac Rx\sinh\frac xR\Big)\,.
\end{aligned}
\end{equation}
We can write a Hamiltonian version of this action with the momenta given by
\begin{equation}
\begin{aligned}
\label{MomentaAdSN1}
    &p_t = M\cosh\frac xR \,,\\
    &p_a = M\left[\frac R{2x} v_a \sinh\frac xR
       +\frac 1{2x^2}x_ax^bv_b\Big(1-\frac Rx\sinh\frac xR\Big)
       -\frac 1{4x^2}x^b \bar\theta\gamma_{ab}\gamma^0\theta \Big(\cosh\frac xR -1\Big)
        \right]\,,\\
    &\bar P_\theta
                 =\frac M2 \bar \theta\gamma^0\,.
\end{aligned}
\end{equation}
Then, the canonical form of \eqref{AdSAction3DN1} is
\begin{equation}
\label{AdSActionCan}
 \hskip -.3truecm   S=\int d\tau \left[ -\dot tE +\dot x_ap^a+\bar {\dot\theta}P_\theta
                 -\frac e2\Big(E^2-M^2\cosh^2 \frac xR\Big )
                 -\Big(\bar P_\theta\cosh\frac xR+\frac12 E \bar\theta\gamma^0\Big )
                   \rho \right]\,.
\end{equation}
The bosonic transformation rules for the coordinates with constant parameters $(\zeta$, $a^i$, $\lambda^i$, $\lambda^i_j)$ corresponding to time translations, spatial translations, boosts and rotations, respectively,  are given by
\begin{equation}
\begin{aligned}
\label{AdSBosTransfN1}
    \delta t&=-\zeta+\frac R{2x} \lambda^k x_k \tanh \frac xR
              +\frac t{Rx}a^k x_k\tanh\frac xR\,,\\[.1truecm]
    \delta x^i&= -\frac 1{x^2}\left(x^ia^k x_k
                 -\frac xR \coth\frac xR (x^ia^kx_k-a^ix^2)\right)
                 -2\lambda^{i}_{\,k}\,x^k \,,\\[.1truecm]
    \delta v^i&= -\lambda^i-\frac1{x^2}\lambda^kx_kx^i\text{sech}\frac xR
     \left(1-\cosh \frac xR\right)
               -2\lambda^{i}_{\,j}\,v^j-\frac{2t}{R^2}a^i\\[.1truecm]
               &\hspace{.4cm}-\frac{2t}{R^2x^2}x^ia^kx_k\,\text{sech}\frac xR\left(1-\cosh\frac xR\right)
               +\frac2{Rx}v_ba^{[i}x^{b]}\text{csch}\frac xR \left( 1-\cosh\frac xR\right)
               \,,\\[.1truecm]
    \delta \theta &= -\frac 12 \lambda^{ab}\gamma_{ab}\theta
                      +\frac1{2Rx}a^kx^b\gamma_{kb}\theta \text{csch}\frac xR
                      \left( 1-\cosh\frac xR\right)\,.
\end{aligned}
\end{equation}
The fermionic transformation rules with constant parameter $\epsilon$ corresponding to the supersymmetry transformation are given by
\begin{equation}
\begin{aligned}
\label{AdSFermTransfN1}
    \delta t&=\frac12\bar\epsilon\gamma^0\theta \text{sech} \frac xR \cosh \frac x{2R}
              -\frac1{2x}x^k\bar\epsilon\gamma^{k0}\theta
                   \text{sech} \frac xR \sinh \frac x{2R}\,,\\[.1truecm]
    \delta x^i&= 0 \,,\\[.1truecm]
    \delta v^i&=\frac 1{Rx}x^i\tanh\frac xR\Big(\bar\epsilon\gamma^0\theta\cosh\frac x{2R}
                 -\frac1x x^k\bar \epsilon \gamma^{k0}\theta\sinh\frac x{2R}\Big) \\
                &-\frac 1{Rx}x^i\bar \epsilon \gamma^0\theta\sinh\frac x{2R}
                 +\frac 1R\bar \epsilon \gamma^{i0}\theta \cosh\frac x{2R}
                 +\frac 1{Rx} x^k\bar \epsilon \gamma^{ik0}\theta\sinh\frac x{2R}\,,\\
    \delta \theta &= \epsilon\cosh\frac x{2R}+\frac1x x^k\gamma_k\epsilon\sinh\frac x{2R}\,.
\end{aligned}
\end{equation}

\subsection{The Super Killing Equations}

The basic Poisson brackets of the canonical variables are given by
\begin{equation}
\begin{aligned}
    &\{E,t\}=1\,,\qquad\{e,\pi_e\}=1\,,\qquad
    \{x_i,p_j\}=\delta_{ij}\,,\\
    &\hspace{.7cm}\{P_\theta^\alpha,\theta_\beta\}=-\delta^\alpha_\beta\,,\qquad
    \{\Pi_\rho^\alpha,\rho_\beta\}=-\delta^\alpha_\beta\,,
\end{aligned}
\end{equation}
and the corresponding Dirac Hamiltonian of the action \eqref{AdSActionCan} is given by
\begin{equation}
    H_D=\frac e2 \Big(E^2-M^2\cosh^2\frac xR\Big)+\lambda \pi_e
              +\Big(\bar P_\theta\cosh\frac xR+\frac12 E \bar\theta\gamma^0\Big )\rho
              +\bar\pi_\rho \Lambda\,,
\end{equation}
$\pi_e=0$ and $\Pi_\rho=0$ are the primary constraints,
$\lambda=\lambda(\tau)$ and $\Lambda=\Lambda(\tau)$ are arbitrary functions.
The corresponding  primary hamiltonian equations of motion are given by
\begin{equation}
\begin{aligned}
\label{AdSeqs1}
    &\dot t=-eE-\frac12\bar \theta \gamma^0\rho\,,\hspace{.5cm}
    \dot x^i=0\,,\hspace{.5cm}
    \dot E=0\,,\hspace{.5cm}
    \dot p^i=\frac {eM^2}{xR}x^i\cosh\frac xR\sinh \frac xR
              -\frac 1{xR}x^i\sinh\frac xR\bar P_\theta\rho  \,,\\
    &\hspace{3.6cm}\dot\pi_e=-\frac12\left(E^2-M^2\cosh^2\frac xR\right)\,,\hspace{.5cm}
    \dot e =\lambda\,,\\
    &\hspace{1.7cm}\dot \theta=-\rho\cosh\frac xR\,,\hspace{.5cm}
    \bar{\dot P}_\theta=-\frac 12 E \bar\rho\gamma^0\,,\hspace{.5cm}
    \dot \rho=-\Lambda\,,\hspace{.5cm}
    \bar{\dot\Pi}_\rho=\bar P_\theta\cosh\frac xR+\frac12E\bar\theta\gamma^0\,.
\end{aligned}
\end{equation}
The stability of primary constraints give as secondary constraint the mass-shell condition $E^2-M^2\cosh^2\frac xR=0$ and the fermionic constraint $\bar P_\theta\cosh\frac xR+\frac12 E \bar\theta\gamma^0=0$. If we require the stability of the secondary constraints we get $\rho=0$. Substituting this into \eqref{AdSeqs1} and using the canonical momenta \eqref{MomentaAdSN1} we obtain equations \eqref{AdSN1eqsmov}.

The generator of canonical transformations has a bosonic and a fermionic part given by
\begin{equation}
    G=-E\xi^0(t,\vec x, \theta)+p_i\,\xi^i(t,\vec x, \theta)+\gamma(t,\vec x, \theta)\pi_e
         -\bar P_\theta\chi(t,\vec x,\theta)+\bar\Pi_\rho\Gamma(t, \vec x, \theta)\,,
\end{equation}
the parameters $\xi^0=\xi^0(t, \vec x, \theta)$, $\xi^i=\xi^i(t, \vec x,\theta)$,
$\chi=\chi(t,\vec x, \theta)$, $\gamma=\gamma(t,\vec x,\theta)$
have the following restrictions
\begin{equation}
\begin{aligned}
 0&= \dot G   \\[.1truecm]
  &= -E(\dot t \partial_t\xi^0+\partial_\theta\xi^0\dot\theta)
               +\dot p_i\xi^i
               + p_i (\dot t \partial_t\xi^i+\partial_\theta\xi^i\dot \theta)+\gamma\dot\pi_e
               -\bar{\dot P}_\theta\chi
               -\bar P_\theta (\partial_t\chi\dot t+\partial_\theta\chi\dot\theta)
               +\bar{\dot \Pi}_\rho\Gamma\\[.1truecm]
               &=eE^2\partial_t\xi^0
               +\frac12 E\partial_t\xi^0\bar\theta\gamma^0\rho
               +E\partial_\theta\xi^0\rho\cosh \frac xR
               +\frac {eM^2}{xR}x^i\xi_i\cosh\frac xR\sinh \frac xR \\[.1truecm]
               &\hspace{.4cm}-\frac 1{xR}x^i\xi_i\sinh\frac xR\bar P_\theta\rho
               -eEp_i\partial_t\xi^i
               -\frac12 p_i\partial_t\xi^i\bar\theta\gamma^0\rho
               -p_i\partial_\theta\xi^i\rho\cosh\frac xR\\[.1truecm]
               & \hspace{.4cm}-\frac12\gamma\left(E^2-M^2\cosh^2\frac xR\right)
               +\frac E2 \bar\rho \gamma^0\chi
               +eE\bar P_\theta\partial_t\chi
               +\frac12 \bar P_\theta\partial_t\chi \bar \theta \gamma^0\rho
               +\bar P_\theta \partial_\theta \chi\rho\cosh\frac xR\\[.1truecm]
               & \hspace{.4cm}+\bar P_\theta \Gamma \cosh \frac xR
               +\frac E2\bar \theta \gamma^0\Gamma\,.
\end{aligned}
\end{equation}
From this equation we derive the super-Killing equations
\begin{equation}
\begin{aligned}
\label{KillingEqsAdSN1}
    &\hspace{1.6cm}\gamma=2e\partial_t \xi^0\,,\hspace{.5cm}
    \Gamma=-\partial_\theta\chi\rho+\frac 1{xR}x^i\xi_i\tanh \frac xR \rho\,,\hspace{.5cm}\\[.1truecm]
    &\partial_t\xi^0=-\frac 1{xR}x^i\xi_i\tanh \frac xR\,,\hspace{.45cm}
    \partial_\theta\xi^0=\frac 12\bar \chi\gamma^0 \text{sech}\frac xR
            +\frac 12\bar \theta \gamma^0\partial_\theta\chi\text{sech}\frac xR\,,\\[.1truecm]
     &\hspace{2.6cm}\partial_t\xi^i=0\,,\hspace{.45cm}
       \partial_\theta\xi^i=0\,,\hspace{.45 cm}
         \partial_t\chi=0\,.\hspace{.45 cm}
\end{aligned}
\end{equation}
The solution to this equations is given by eqs.~\eqref{AdSBosTransfN1} and \eqref{AdSFermTransfN1} with the symmetry generator $G$ given by
\begin{equation}
\begin{aligned}
    G&=-E\xi^0(\vec x, \theta)+p_i\,\xi^i(t,\vec x)
        +2e\partial_t \xi^0(\vec x, \theta)\pi_e
         -\bar P_\theta\chi(\vec x,\theta)\\[.1truecm]
         &\hspace{.4cm}+\bar\Pi_\rho\Big(-\partial_\theta\chi(\vec x, \theta)\rho
                 +\frac 1{xR}x^i\xi_i(\vec x, \theta)\tanh \frac xR \rho\Big)\,.
\end{aligned}
\end{equation}
Then, the $\mathcal N=1$ AC superparticle has an infinite dimensional algebra with the transformation rules given by \eqref{AdSBosTransfN1} and \eqref{AdSFermTransfN1}.

\subsection{The Flat Limit}

We end this section with some comments on the flat limit ($R\rightarrow\infty$) which can be taken directly from the AC curved case in order to obtain the dynamics and symmetries of the $\mathcal N=1$ flat Carroll superparticle. In this case, the time and space components of the supervielbein simplify to
\begin{equation}
\begin{aligned}
   E^0 &=dt-\frac12\bar{\theta}\gamma^0d\theta,\,\hskip 1truecm
   E^a&=dx^a\,,\hskip 1truecm
    E_\alpha&= d\theta_\alpha\,.
\end{aligned}
\end{equation}
In the $R\rightarrow\infty$ limit, the torsion becomes $T_0=-\frac12d\bar{\theta}\gamma^0d\theta$ and since we are studying the flat case, the spin connection vanishes. The supertranslations can be again written in terms of the supervielbein in matrix form as in \eqref{matrixsupertrans} and the action is given by
\begin{equation}
\label{FlatBosAct}
    S= M\int \Big(E^0+\frac 12 v_a  E^a\Big)^*=M\int d\tau(\dot t-\frac12\bar{\theta}\gamma^0\dot\theta
       +\frac12 v_a \dot x^a)\,.
\end{equation}
The equations of motion that follow from this action are: 
\begin{equation}
    \dot{\vec x} = \dot{\vec v} = \dot{\theta}=0\,.
\end{equation}
Therefore, the superparticle does not move. The transformation rules of the different variables are given by
\begin{equation}
\begin{aligned}
\label{N1Transf}
    \delta t&=-\zeta+\frac12\lambda^i x_i +\frac12\bar\epsilon\gamma^0\theta\,,\hspace{2cm}
    \delta x^i= -a^i -2\lambda^{i}_{\,j}\,x^j\,,\\
    \delta v^i&= -\lambda^i-2\lambda^{i}_{\,j}\,v^j \,,\hspace{3.15cm}
    \delta \theta = -\frac 12 \lambda_{ij}\gamma^{ij}\theta+\epsilon\,.
\end{aligned}
\end{equation}
As we can see from the transformation of $\theta$ the $\mathcal N=1$ Carroll superparticle is not BPS like in the relativistic and Galilean case.

If we rewrite the action \eqref{FlatBosAct} in  Hamiltonian form
\begin{equation}
    S=\int d\tau \left[ -\dot tE +\dot x_ap^a+\bar {\dot\theta}P_\theta
                     -\frac e2(E^2-M^2 )
                     -\Big(\bar P_\theta+\frac12 E \bar\theta\gamma^0\Big )
                       \rho  \right]\,,
\end{equation}
it turns out that the super-Killing equations can be obtained as the flat limit of the equations \eqref{KillingEqsAdSN1}
\begin{equation}
\begin{aligned}
    &\gamma=0\,,\hspace{.5cm}
    \Gamma=-\partial_\theta\chi\rho\,,\hspace{.5cm}
    \partial_t\xi^0=0\,,\hspace{.5cm}
    \partial_t\xi^i=0\,,\hspace{.5cm}
    \partial_\theta\xi^i=0\,,\hspace{.5 cm}
    \partial_t\chi=0\,,\\
    &\hspace{4.5cm} \partial_\theta\xi^0=\frac 12\bar \chi\gamma^0
            +\frac 12\bar \theta \gamma^0\partial_\theta\chi\,,
\end{aligned}
\end{equation}
where the symmetry generator $G$ is
\begin{equation}
    G=-E\xi^0(\vec x, \theta)+p_i\,\xi^i(\vec x)
         -\bar P_\theta\chi(\vec x,\theta)
         -\bar\Pi_\rho\partial_\theta\chi(\vec x,\theta)\rho\,.
\end{equation}
From the variation of the momenta
\begin{equation}
    \delta p_i=\{p_i,G\}
         =E\partial_i\xi^0
          -p_k\partial_i\xi^k+\bar P_\theta\partial_i\chi\,
\end{equation}
and using that the energy, the spatial momenta and the fermionic momenta are given by
\begin{equation}
    \label{symPflat}
    E=-M\,,\qquad
    p_i=\frac M2 v_i\,,\qquad
    \bar P_\theta =\frac M2 \bar{\theta} \gamma^0\,,
\end{equation}
we find that the transformation rule of $v^i$
\begin{equation}
    \delta v_i
         =-2\partial_i\xi^0
          - v_k\partial_i\xi^k+ \bar \theta \gamma^0\partial_i\chi\,.
\end{equation}

Note that the above  symmetries include  the dilatations given by
\begin{equation}
    \delta t=0,\,\hskip 1truecm  \delta x^a= x^a,\,\hskip 1truecm  \delta\theta=0,\,\hskip 1truecm \delta v^a=-v^a\,.
    \end{equation}
These dilatations, together with the super-Carroll transformations, form a supersymmetric extension of the Lifshitz Carroll algebra  \cite{Gibbons:2009me}  with dynamical exponent z=0. The Lifshitz Carroll algebra with z=0 has appeared in a recent study of warped conformal field theories \cite{Hofman:2014loa}.

\section{The \texorpdfstring{$\mathcal N=2$}{N=2} Flat Carroll Superparticle}

In this Section we extend our investigations to the ${\cal N}=2$ supersymmetric case. The flat case is discussed in this Section while the curved case will be dealt with in Appendix C.

\subsection{The \texorpdfstring{$\mathcal N=2$}{N=2} Carroll Superalgebra}

Our starting point is the $\mathcal N =2$ super-Poincar\'e algebra. For simplicity, we consider 3D only. The basic commutators are $(A=0,1,2;i=1,2)$
\begin{equation}
\begin{aligned}
    \left[M_{AB}, M_{CD}\right]&=2\eta_{A[C}M_{D]B}-2\eta_{B[C}M_{D]A}\,,\\
    [M_{AB},P_C]&=2\eta_{C[B}P_{A]}\,,\\
    [M_{AB},Q^i]&=-\frac12\gamma_{AB}Q^i\,,\\
    \{Q^i_\alpha,Q^j_\beta\}&=2[\gamma^AC^{-1}]_{\alpha\beta}P_A\delta^{ij}
                              +2[C^{-1}]_{\alpha\beta}\epsilon^{ij}Z\,.
\end{aligned}
\end{equation}
To make the Carroll contraction we define new supersymmetry charges by
\begin{equation}
    Q_\alpha^\pm=\frac12(Q_\alpha^1\pm\gamma_0Q_\alpha^2)
\end{equation}
and rescale the different symmetry generators with a parameter $\omega$ as follows:
\begin{equation}
\begin{aligned}
    P_0=\frac \omega2 H\,,\qquad
    M_{a0}&=\omega K_a\,,\qquad
    Z=\omega \tilde Z\,,\\
    Q^+=\sqrt{\omega}\tilde Q^+\,,&\qquad Q^-=\sqrt{\omega} \tilde Q^-\,.
\end{aligned}
\end{equation}
Taking the limit $\omega\to\infty$ we obtain the following 3D Carroll algebra
\begin{equation}
\begin{aligned}
\label{3DN2C1_alg}
    \left [M_{ab}, K_c\right]&=2\delta_{c[b}K_{a]}\,,\hspace{3.2cm}
    [M_{ab}, P_c]=2\delta_{c[b}P_{a]}\,,\hspace{1.75cm}\\[.1truecm]
    \left[K_a, P_b\right]&=-\frac12\delta_{ab}H\,,\hspace{3cm}
    [M_{ab},\tilde Q^\pm]=-\frac12\gamma_{ab}\tilde Q^\pm\,\\[.1truecm]
    \{\tilde Q^+_\alpha,\tilde Q^+_\beta\}&=[\gamma^0C^{-1}]_{\alpha\beta}
                                            \Big(\frac12 H+\tilde Z\Big)\,,
    \hspace{.85cm}
    \{\tilde Q^-_\alpha,\tilde Q^-_\beta\}= [\gamma^0C^{-1}]_{\alpha\beta}
                                            \Big(\frac12 H-\tilde Z\Big)\,.
\end{aligned}
\end{equation}

The Maurer-Cartan equation $dL^C-\frac 12 f^C_{AB} L^B\wedge L^A=0$ in components reads:
\begin{equation}
\begin{aligned}
    dL_H&= -\frac12 L_P^{\;a}\,L_K^{\;a}-\frac14 \bar L_-\gamma^0L_-
           -\frac14 \bar L_+\gamma^0L_+ \,,\hspace{1.5cm}
    d L^{\;a}_P= 2L_P^{\;b}\,L_M^{\;ab}\,,\\[.1truecm]
    d L_Z&=-\frac12\bar L_+\gamma^0L_++\frac12\bar L_-\gamma^0L_-\,,\hspace{3.4cm}
    d L^{\;a}_K= 2L_K^{\;b}\,L_M^{\;ab}\,,\\[.1truecm]
    d L_-&=\frac12 \gamma_{ab} L_- \,L_M^{\;ab}\,,\hspace{5.6cm}
    d L_+=\frac12 \gamma_{ab} L_+ \,L_M^{\;ab}\,,\\[.1truecm]
    d L_M^{\;ab}&=2L_M^{\;ca}\,L_M^{\;cb}\,.
\end{aligned}
\end{equation}

\subsection{Superparticle Action and Kappa Symmetry}

To construct the action of the $\mathcal N=2$ Carrollian superparticle we consider the following coset:
\begin{equation}
 \frac G H=\frac {\mathcal N=2\textrm { super Carroll}}{\textrm {SO(D-1)}}\,.
\end{equation} 
The coset element is given by $g= g_0\,U,$ where $g_0=e^{Ht} e^{ P_a x^a} e^{Q_\alpha^-\theta_-^\alpha} e^{Q_\alpha^+\theta_+^\alpha}e^{Zs}$ is the coset representing  the `empty' $\mathcal N=2$ Carroll superspace with a central charge extension and $U=e^{ K_a v^a}$ is a general Carroll boost representing the insertion of the particle.

The Maurer-Cartan form associated to the super-Carroll space is given  by
\begin{equation}
\Omega_0=(g_0)^{-1}dg_0=H E^0+P_aE^a - \bar Q^- E_- - \bar Q^+ E_+ + Z E_Z\,,
\end{equation}
where $(E^0, E^a, {E_-}_{\alpha},{E_+}_{\alpha},E_Z )$ are the supervielbein components of the Carroll superspace given explicitly by
\begin{equation}
\begin{aligned}
    E^0& = d t
           - \frac14 \bar\theta_-\gamma^0 d\theta_-
           - \frac14 \bar\theta_+\gamma^0 d\theta_+ \,, \hspace{1.5cm}
    E^a =dx^a \,,\\
    {E_-}_\alpha&=d{\theta_-}_\alpha \,,\hspace{5.2cm}
    {E_+}_\alpha=d{\theta_+}_\alpha\,,\\
    E_Z &= ds
                + \frac12 \bar \theta_-\gamma^0d\theta_-
                - \frac12 \bar \theta_+\gamma^0d\theta_+\,.\hspace{3.15cm}
\end{aligned}
\end{equation}
In terms of the supervielbein  the Maurer-Cartan form of the $\mathcal N=2$ Carroll superparticle is given by
\begin{equation}
\begin{aligned}
\label{3DLs}
    L_H& = E^0
           + \frac12 v_aE^a \,, \hspace{1.5cm}
    L_P^a =E^a \,,\\[.1truecm]
    L_Z &= E_Z\,,\hspace{3.15cm}
    L^a_K = dv^a\,,\\[.1truecm]
    {L_-}_\alpha&={E_-}_\alpha \,,\hspace{2.8cm}
    {L_+}_\alpha={E_+}_\alpha\,.
\end{aligned}
\end{equation}
As before, we can write  the space-time super-translations in matrix form in terms of the Vielbein of Carroll superspace as follows:
\begin{equation}
  \big (\arraycolsep=4pt \begin{array}{ccccc} L_H, & {L_P}^a,&  {L_-}_{\alpha},& {L_+}_{\alpha}, & L_Z \end{array} \big)=  \big ( \begin{array}{ccccc} E^0, & E^a, & {E_-}_{\alpha},& {E_+}_{\alpha}, & E_Z\end{array} \big) \begin{pmatrix} 1 & 0 & 0 & 0 & 0\\ \frac12 v_a & 1& 0& 0 & 0\\ 0 & 0 &1 & 0& 0 \\  0 & 0 & 0 & 1& 0\\  0& 0& 0& 0& 1\end{pmatrix}\,.
\end{equation}

The action of the ${\cal N}=2$ Carrollian superparticle is given by the pull-back of all the $L$'s that are invariant under rotations:
\begin{equation}
\label{Action3D}
\begin{aligned}
    S&=a\int(L_H)^*+b\int(L_Z)^* \\
     &=a\int d\tau \left( \dot t
           - \frac14 \bar\theta_-\gamma^0 \dot\theta_-
           - \frac14 \bar\theta_+\gamma^0 \dot\theta_+
           + \frac12 v_a\dot x^a \right)
      +b\int d\tau \left( \dot s
           +\frac12\bar \theta_-\gamma^0\dot\theta_-
           -\frac12\bar \theta_+\gamma^0\dot\theta_+\right) \,.
\end{aligned}
\end{equation}
The equations of motion corresponding to this action are given by
\begin{equation}
    \dot x_a = 0\,,\quad
    \dot v_a=0\,,\quad
    \dot \theta_- =0\,,\quad
    \dot \theta_+=0\,.
\end{equation}
The transformation rules for the coordinates with constant parameters
$(\zeta$, $\eta$,   $a^i$, $\lambda^i$, $\lambda^i_j$, $\epsilon_+$, $\epsilon_-)$
corresponding to time translations, $Z$ transformations, spatial translations, boosts, rotations and supersymmetry transformations, respectively, are given by
\begin{equation}
\begin{aligned}
    \delta t&=-\zeta+\frac12 \lambda^i x_i
              +\frac14\bar\epsilon_-\gamma^0\theta_-
              +\frac14\bar\epsilon_+\gamma^0\theta_+\,,\hspace{1.5cm}
    \delta x^i= -a^i
                 -2\lambda^{i}_{\,j}\,x^j \,,\\
    \delta s &=-\eta
               -\frac12\bar{\epsilon}_-\gamma^0\theta_-
               +\frac12\bar{\epsilon}_+\gamma^0\theta_+ \,,\hspace{3cm}
    \delta v^i= -\lambda^i-2\lambda^{i}_{\,j}\,v^j \,,\\
    \delta \theta_+ &= -\frac 12 \lambda^{ab}\gamma_{ab}\theta_++\epsilon_+\,,\hspace{4.45cm}
    \delta \theta_- = -\frac 12 \lambda^{ab}\gamma_{ab}\theta_-+\epsilon_-\,.\\
\end{aligned}
\end{equation}

To derive an action that is invariant under additional $\kappa$-transformations we need to find a fermionic gauge-transformation that leaves $L_H$ and/or $L_Z$ invariant. The variation of $L_H$ and $L_Z$ under gauge-transformations is given by
\begin{equation}
\label{gaugesymflat}
\begin{aligned}
   \delta L_H&=d([\delta z_H])+\frac12L_P^a[\delta z_K^a]+\frac12L_K^a[\delta z_P^a]
               +\frac12\bar L_-\gamma^0[\delta z_-]
               +\frac12 \bar L_+\gamma^0[\delta z_+]\,,\\[.1truecm]
   \delta L_Z&= d([\delta z_Z])-\bar L_-\gamma^0[\delta z_-]+\bar L_+\gamma^0[\delta z_+]\,.
\end{aligned}
\end{equation}
where $[\delta z_K^a]$ is obtained from $L_H$ by changing the 1-forms $dt,$ $ d\theta_+,$  $d\theta_-$ with the transformations $\delta t,$ $\delta\theta_+,$  $\delta\theta_-$. In analogous way we can construct the other terms appearing in \eqref{gaugesymflat}.

For $\kappa$-transformations, $[\delta z_H]=0,$ $[\delta z_K^a]=0,$ $[\delta z_P^a]=0$,
\begin{equation}
\begin{aligned}
    0=\delta L_H&=\frac12\delta\bar \theta_-\gamma^0[\delta z_-]
                    +\frac12\delta\bar \theta_+\gamma^0[\delta z_+] \,,\\
    0=\delta L_Z&=-\delta\bar \theta_-\gamma^0[\delta z_-]
                    +\delta\bar \theta_+\gamma^0[\delta z_+]\,.
                    \end{aligned}
\end{equation}
It follows that to obtain a $\kappa$-symmetric action we need to take
$b=\pm\frac12a$. We focus here on the case
$b=-\frac12 a$. With this choice  the action and $\kappa$-symmetry rules are given by
\begin{equation}
\begin{aligned}
    S&=a\int (L_H-\frac12 L_Z)^*\,, \qquad [\delta z_+]=\kappa\,,\qquad [\delta z_-]=0\,,
\end{aligned}
\end{equation}
\noindent where $\kappa=\kappa(\tau)$ is an arbitrary local parameter. Using this we find the following $\kappa$-transformations of the coordinates
\begin{equation}
\begin{aligned}
    \delta t&=\frac14\bar \theta_+\gamma^0\kappa\,,\hspace{1.3cm}
    \delta x^a=0\,,\hspace{1.3cm}
    \delta\theta_+=\kappa\,,\\
    \delta s&=\frac12\bar \theta_+\gamma^0\kappa\,,\hspace{1.35cm}
    \delta v_a =0\,,\hspace{1.3cm}
    \delta \theta_-=0\,.
\end{aligned}
\end{equation}
After fixing the $\kappa$-symmetry, by imposing the gauge condition $\theta_+=0$, the action reduces to
\begin{equation}
\label{Action3Dkfixflat}
\begin{aligned}
    S&=a\int d\tau \left( \dot t -\frac12\dot s
           - \frac12 \bar\theta_-\gamma^0 \dot\theta_-
           + \frac12 v_a\dot x^a \right) \,.
\end{aligned}
\end{equation}
The residual transformations that leave this action invariant are given by
\begin{equation}
\begin{aligned}
    \delta t&=-\zeta+\frac12 \lambda^i x_i
              +\frac14\bar\epsilon_-\gamma^0\theta_-\,,\hspace{1.5cm}
    \delta x^i= -a^i
                 -2\lambda^{i}_{\,j}\,x^j \,,\\
    \delta s &=-\eta
               -\frac12\bar{\epsilon}_-\gamma^0\theta_- \,,\hspace{3cm}
    \delta v^i= -\lambda^i-2\lambda^{i}_{\,j}\,v^j \,,\\
    \delta \theta_- &= -\frac 12 \lambda^{ab}\gamma_{ab}\theta_-+\epsilon_-\,.\\
\end{aligned}
\end{equation}
The linearly realized supersymmetry acts trivially on all the fields and therefore the ${\cal N}=2$ Super Carroll particle reduces to the ${\cal N}=1$ Super Carroll particle and hence is not BPS since the kappa-symmetry eliminates the linearized supersymmetry. This is different from the ${\cal N}=2$ Super Galilei case were BPS particles do exist.

\section{Discussion and Outlook}

In this paper we have investigated the geometry of the flat and curved (AdS) Carroll space both in the bosonic as well as in the supersymmetric case. We furthermore have analyzed the symmetries of a particle moving in such a space. In the bosonic case we constructed the Vielbein and  spin connection of the AdS Carroll (AC) space which shows that this space is torsionless with constant (negative) curvature. We constructed the action of a massive particle moving in this space thereby extending the flat case analysis of \cite{Bergshoeff:2014jla}. Like in the flat case, we found that the  AC particle does not move. However, in the curved case the momenta are  not conserved. Particles moving in a  Carroll space, whether  flat or curved, do not have a relation among their velocities and momenta.

Using the symmetries of the AC particle we have computed the Killing equations of the AC space. We found that these Killing equations allow an infinite-dimensional algebra of symmetries that, unlike in the flat case,  does not include dilatations. Another difference with the flat case is that there is no duality between the Newton-Hooke and  AdS Carroll algebras. Furthermore, in the curved case the mass-shell constraint depends on the coordinates of the AC space.

In the second part of this paper we have extended our investigations  to the supersymmetric case. Unlike the bosonic case, the ${\cal N}=1$ AC superspace  has torsion with constant curvature due to the presence of fermions. Like in the bosonic case, we found that the ${\cal N}=1$  AC superparticle does not move and the momenta are conserved.  We have constructed the super-Killing equations and showed that the symmetries form an infinite dimensional superalgebra. After taking the flat limit we found that  among the symmetries of the ${\cal N}=1$ Carroll superparticle we have a supersymmetric extension of the Lifshitz Carroll algebra \cite{Gibbons:2009me} with dynamical exponent $z=0$. The bosonic part of this algebra has appeared as a symmetry  of warped conformal field theories \cite{Hofman:2014loa}.

We also showed that the ${\cal N}=2$ Carroll superparticle has a fermionic kappa-symmetry such that, when this gauge symmetry is fixed, the ${\cal N}=2$ Carroll superparticle reduces to the ${\cal N}=1$ Carroll superparticle. Apparently, in flat Carroll superspace  the number of supersymmetries is not physically relevant. This is due to the fact that the kappa gauge symmetry neutralizes the extra linear supersymmetries beyond ${\cal N}=1$. Unlike the bosonic case, there is no duality between  the ${\cal N}=2$ Super Galilei and Super Carroll algebras.

In a separate appendix we investigated the ${\cal N}=2$ AC superparticle\,\footnote{ For simplicity we did only consider the 3D case.}. We  studied the so-called (2,0) and (1,1) super-Carroll spaces and the corresponding superparticles. Physically, the (2,0) and (1,1) cases are different, they have unequal degrees of freedom.  For instance, only the (2,0) superparticle has a kappa-symmetry.  Apparently, for the  AC superparticle the type of supersymmetry one considers does make a difference.

As  a possible continuation to the ideas presented in this paper it would be interesting to find the coupling of the AdS Carroll particle, and the corresponding superparticle, to the (super) AdS gauge fields. Like in the flat Carroll case \cite{Bergshoeff:2014jla} we expect that the (super) particle will have a non-trivial dynamics.

Finally, it would be interesting  to study if one could construct the corresponding Carroll (super) gravity theory.  There are two approaches to this issue. One approach is to gauge the (super) Carroll algebra and/or the Lifhsitz Carroll algebra with $z=0$. In this respect we note that the gauging of the Carroll algebra as performed in \cite{Bergshoeff:2014jla} can be improved by imposing curvature constraints that allow to set some of the spin-connection fields equal to zero, like in \cite{Hofman:2014loa}, instead of trying to solve for all of the spin-connection fields. It would be interesting to apply this improved gauging technique to the other algebras as well. A second alternative approach would be to try to define an ultra-relativistic limit of relativistic (super-)gravity similar to the non-relativistic limit.

\section*{Acknowledgments}

We acknowledge useful discussions with Blaise Rollier and Hamid Afshar. In particular, we are grateful to  Blaise Rollier for an elucidating explanation of his recent work \cite{Hofman:2014loa}. This work is partially financed by   FPA2013-46570,  2014-SGR-104, CPAN, Consolider CSD 2007-0042. The work of L.P. is supported by an Ubbo Emmius sandwich scholarship from the University of Groningen.

\newpage

\appendix

\section{The Carroll action as a Limit of the AdS action}

In this appendix we show how  to obtain the action of the $D$-dimensional free AdS Carroll particle starting from the massive particle moving in an $D$-dimensional  AdS spacetime and to take the Carroll limit. The canonical form of the action before taking the limit is given by
\begin{equation}
    S=\int d\tau[p^\mu \dot x_\mu -\frac {\tilde e}2(g_{\mu\nu}p^\mu p^\nu+m^2)]\,,
\end{equation}
where $\tau$ is the evolution parameter, $g_{\mu\nu}$ is the metric of an AdS space and $\tilde e$ is a Lagrange multiplier. We use that the signature of the metric is $(-,+,+,\ldots)$ and that the AdS line element is given by
\begin{equation}
    ds^2=-\cosh^2\frac xR (dx^0)^2+\frac {R^2}{x^2}\sinh^2\frac xR(dx^a)^2
    -\left( \frac {R^2}{x^2} \sinh^2\frac xR -1\right)(dx)^2\,,
\end{equation}
where $x=\sqrt{x_ax^a}$. To take the Carroll limit we first consider a re-scaling of the variables
\begin{equation}
    x^0=\frac t\omega\,,\quad
    p^0=\omega E\,,\quad
    m=\omega M\,,\quad
    \tilde e=-\frac e{\omega^2}\,,
\end{equation}
and next take the limit  $\omega\rightarrow\infty$ to obtain
\begin{equation}
    S=\int d\tau[-E\dot t+p^a\dot x_a-\frac e2(E^2-M^2\cosh^2\frac xR) ]\,.
\end{equation}
The equations of motion are given by
\begin{equation}
\begin{aligned}
\label{ecsmovAdSLim}
    \dot t &=-eE\,,\hspace{.5cm} \dot E=0\,,\\[.1truecm]
    \dot x^a&=0\,,\hspace{1.05cm} \dot p^a=\frac{eM^2}{Rx}x^a \cosh\frac xR\sinh\frac xR\,,\\[.1truecm]
    \dot e&=\lambda\,,\hspace{1.05cm}\pi_e=-\frac 12(E^2-M^2\cosh^2\frac xR)\,.
\end{aligned}
\end{equation}
Note that although the dynamics of $x$ is trivial, i.e.~$\dot x^a=0$ (the particle is not changing its position), the momentum is changing over $\tau$ because $\dot p^a \neq0$. In the flat limit (the limit when $R\rightarrow\infty$) the particle is at rest and does not move.

Finally, the mass-shell constraint reads
\begin{equation}
E^2-M^2\cosh^2\frac xR=0\,.
\end{equation}

\section{The super-AdS Carroll action as a limit of the super-AdS action}

In the supersymmetric case, we obtain the action of the free AdS Carroll superparticle starting from the massive superparticle moving in an AdS spacetime whose action is given by
\begin{equation}
    S=\int d\tau[\dot x_\mu p^\mu + \bar { \dot \phi } P_\phi
                    -\frac {\tilde e}2(g_{\mu\nu}p^\mu p^\nu+m^2)
                    +(\bar P_\phi +g_{\mu\nu} p^\mu \bar \phi \gamma^\nu) \lambda ]\,,
\end{equation}
where $g_{\mu\nu}$ is the AdS metric with line element given by eq.~(A.2). Rescaling the variables as
\begin{equation}
\begin{aligned}
    &x^0=\frac t\omega\,,\quad
    p^0=\omega E\,,\quad
    m=\omega M\,,\quad
    \tilde e=-\frac e{\omega^2}\,,\\
    &\hspace{.7cm}\phi = \frac 1{\sqrt \omega}\theta\,,\quad
    P_\phi = \sqrt{\omega}\,P_\theta \,,\quad
    \lambda=\frac 1{\sqrt \omega} \rho\,,
\end{aligned}
\end{equation}
allows us to take the Carroll limit with $\omega\rightarrow\infty$ to obtain
\begin{equation}
    S=\int d\tau[-\dot tE+\dot x_ap^a + \bar{\dot \theta}P_\theta
       -\frac e2\left (E^2-M^2\cosh^2\frac xR\right )
       +(\bar P_\theta\cosh\frac xR + E\, \bar \theta \gamma^0) \rho  ]\,.
\end{equation}
The primary equations of motion are
\begin{equation}
\begin{aligned}
\label{ecsmovAdSLimN1}
    \dot t &=-eE-\bar \theta \gamma^0\rho \,,\hspace{.5cm} \dot E=0\,,\\
    \dot x^a&=0\,,\hspace{2.3cm}
    \dot p^a=\frac{eM^2}{Rx}x^a \cosh\frac xR\sinh\frac xR
              -\frac1{xR}x^a\sinh\frac xR\bar P_\theta\rho \,,\\
    \dot e&=\lambda\,,\hspace{2.3cm}
    \pi_e=-\frac 12(E^2-M^2\cosh^2\frac xR)\,,\\
    \dot \theta &=-\cosh\frac xR \rho \,,\hspace{.75cm}
    \bar{\dot P}_\theta = -E\,\bar \rho\gamma^0\,,\\
    \dot \rho&=-\Lambda \,,\hspace{1.9cm}
    \bar {\dot {\Pi}}_\rho = \bar P_\theta\cosh\frac xR+E\,\bar \theta\gamma^0\,.
\end{aligned}
\end{equation}
After requiring the stability of all the constraints we obtain the equations of motion \eqref{AdSN1eqsmov}. Like in the bosonic case we find that the dynamics of $x$ is trivial, $\dot x^a=0$ (the particle is not changing its position), but that the momentum is changing over $\tau$ because $\dot p^a \neq0$.

\section{The 3D \texorpdfstring{$\mathcal N=2 $}{N=2} AdS Carroll Superparticle}

There are two independent versions of the 3D ${\cal N}=2$ AdS algebra, the so-called $\mathcal{N}=(1,1)$ and $\mathcal{N}=(2,0)$ algebras. Correspondingly, there are two  possible ${\cal N}=2$ AdS Carroll superalgebras which we consider below.

\subsection{The  \texorpdfstring{$\mathcal N=(2,0) $}{N=(2,0)} AdS Carroll Superalgebra}

We will start with the contraction of the 3D $\mathcal N=(2,0)$ AdS algebra.
The basic commutators are given by $(A=0,1,2; i=1,2)$
\begin{equation}
\begin{aligned}
    \left[M_{AB}, M_{CD}\right]&=2\eta_{A[C}M_{D]B}-2\eta_{B[C}M_{D]A}\,,\hspace{1.5cm}
    [M_{AB},Q^i]=-\frac12\gamma_{AB}Q^i\,,\\[.1truecm]
    [M_{AB},P_C]&=2\eta_{C[B}P_{A]}\,,\hspace{4.7cm}
    [P_A,Q^i]=x\gamma_A Q^i\,,\\[.1truecm]
    [P_A,P_B]&=4x^2M_{AB}\,,\hspace{4.95cm}
    [\mathcal R, Q^i]=2x\epsilon^{ij}Q^j\,,\\[.1truecm]
    \{Q^i_\alpha,Q^j_\beta\}&=2[\gamma^AC^{-1}]_{\alpha\beta}P_A \delta^{ij}
                              +2x[\gamma^{AB}C^{-1}]_{\alpha\beta}M_{AB}\delta^{ij}
                              +2[C^{-1}]_{\alpha\beta}\epsilon^{ij}\mathcal R\,.
\end{aligned}
\end{equation}
Here $P_A,M_{AB},\mathcal R$ and $Q^i_\alpha$ are the generators of space-time translations, Lorentz rotations, SO(2) R-symmetry transformations and supersymmetry transformations, respectively. The bosonic generators $P_A, M_{AB}$ and $\mathcal R$ are anti-hermitian while de fermionic generators $Q^i_\alpha$ are hermitian. The parameter $x=1/(2R)$, with $R$ being the AdS radius. Note that the generator of the SO(2) R-symmetry becomes the central element of the Poincar\'e algebra in the flat limit $x\rightarrow0$.

To take the Carroll contraction we define new supersymmetry charges by
\begin{equation}
Q^\pm_\alpha=\frac 12(Q^1_\alpha\pm\gamma_0Q^2_\alpha)
\end{equation}
and rescale the generators with a parameter $\omega$ as follows:
\begin{equation}
\begin{aligned}
    P_0&=\frac\omega2 H\,,\hspace{1cm}
    \mathcal R = \omega Z\,, \hspace {1cm}
    M_{a0}=\omega K_a\,,\hspace{1cm}
    Q^\pm&=\sqrt\omega\tilde Q^\pm\,.
\end{aligned}
\end{equation}
Taking the limit $\omega\rightarrow\infty$ and dropping the tildes on the $Q^\pm$ we get the following 3D $\mathcal N=(2,0)$ Carroll superalgebra:
\begin{equation}
\begin{aligned}
    \label{AdSCarrAlg}
    \left[M_{ab},P_c\right]&=2\delta_{c[b}P_{a]}\,, \hspace{1cm}
    [M_{ab},K_c]=2\delta_{c[b}K_{a]}\,, \\[.1truecm]
    [P_a, P_b]&=\frac1{R^2}M_{ab}\,,\hspace{1cm}
    [P_a,K_b]=\frac12\delta_{ab}H\,,\hspace{1cm}
    [P_a,H]=\frac 2{R^2}K_a\\[.1truecm]
    [P_a,Q^\pm]&=\frac 1{2R}\gamma_a Q^\mp\,,\hspace{1cm}
    [M_{ab},Q^\pm]=-\frac 12 \gamma_{ab}Q^\pm\,,\\[.1truecm]
    \{ Q^+_\alpha, Q^+_\beta\}&=\frac12[\gamma^0C^{-1}]_{\alpha\beta}
                                            \left(H+ 2Z\right)\,,\hspace{.55cm}
    \{ Q^-_\alpha, Q^-_\beta\}= \frac 12[\gamma^0C^{-1}]_{\alpha\beta}
                                            \left( H-2 Z\right)\,,\\[.1truecm]
    \{ Q^+_\alpha, Q^-_\beta\}&=\frac 1R [\gamma^{a0}C^{-1}]_{\alpha\beta}K_a\,.
\end{aligned}
\end{equation}
In components the Maurer-Cartan equation  $dL^C-\frac12\,{f^C}_{AB}L^BL^A=0$ reads as follows:
\begin{equation}
\begin{aligned}
    dL_H=& -\frac12 L_P^{\;a}\,L_K^{\;a}-\frac14 \bar L_-\gamma^0L_-
           -\frac14 \bar L_+\gamma^0L_+ \,,\hspace{1cm}
    d L^{\;a}_P= 2L_P^{\;b}\,L_M^{\;ab}\,,\\[.1truecm]
    d L^{\;a}_K=& 2L_K^{\;b}\,L_M^{\;ab}+\frac2{R^2}L_H{L_P}^a
                  -\frac1R\bar L_-\gamma^{a0}L_+\,,\hspace{1.15cm}
    d L_Z=-\frac12\bar L_+\gamma^0L_+ +\frac12\bar L_-\gamma^0L_-\,, \\[.1truecm]
    d L_-=&\frac12 \gamma_{ab} L_- \,L_M^{\;ab}-\frac1{2R}\gamma_aL_+{L_P}^a\,,\hspace{2.7cm}
    d L_+=\frac12 \gamma_{ab} L_+ \,L_M^{\;ab}-\frac1{2R}\gamma_aL_-{L_P}^a\,,\\[.1truecm]
    d L_M^{\;ab}=&2L_M^{\;ca}\,L_M^{\;cb}+\frac1{2R^2}{L_P}^b{L_P}^a\,.
\end{aligned}
\end{equation}

\subsection{Superparticle action}

We use the algebra \eqref{AdSCarrAlg} to construct the action of the $\mathcal N=2$ Carrollian superparticle. The coset that we will consider  is
\begin{equation}
 \frac G H=\frac {\mathcal N=(2,0)\textrm { AdS Carroll}}{\textrm {SO(D-1)}}\,,
\end{equation} 
with the coset element $g= g_0\,U,$ where $g_0=e^{Ht} e^{ P_a x^a} e^{Q_\alpha^-\theta_-^\alpha} e^{Q_\alpha^+\theta_+^\alpha}e^{Zs}$ is the coset representing  the $\mathcal N=(2,0)$ Carroll superspace with a central charge extension and $U=e^{ K_a v^a}$ is a general Carroll boost that represents the superparticle.

The Maurer-Cartan form associated to the super-Carroll space is given by
\begin{equation}
\Omega_0=(g_0)^{-1}dg_0=H E^0+P_aE^a + K_a \omega^{a0}+
M_{ab}\omega^{ab} - \bar Q^- E_- - \bar Q^+ E_+ + Z E_Z\,,
\end{equation}
where $(E^0, E^a, {E_-}_{\alpha},{E_+}_{\alpha},E_Z )$ and $(\omega^{a0},\omega^{ab})$ are the supervielbein and the spin connection of the Carroll superspace which are given explicitly by
\begin{equation}
\begin{aligned}
    E^0& =  d t \cosh\frac xR
           - \frac14 (\bar\theta_-\gamma^0 d\theta_-
           +  \bar\theta_+\gamma^0 d\theta_+)
           -\frac 14 \omega^{ab} (\bar\theta_+\gamma_{ab}\gamma^0\theta_+
                                      +\bar\theta_-\gamma_{ab}\gamma^0\theta_-)\\[.1truecm]
             &\hspace{.5cm} +\frac1{4R}\bar\theta_-\gamma^{a0}\theta_+ E^a \,,\\[.1truecm]
    E^a &=\frac Rxdx^a\sinh\frac xR
                  +\frac 1{x^2}x^ax^bdx_b\Big(1-\frac Rx \sinh \frac xR\Big) \,,\\[.1truecm]
    {\omega}^{a0} &= -\frac 2{xR}dtx^a\sinh\frac xR
                   -\frac 1R \bar\theta_+\gamma^{a0}d\theta_-
                   -\frac1{4R^2}(\bar\theta_+\gamma_{ab}\gamma^0\theta_+
            +\bar\theta_-\gamma_{ab}\gamma^0\theta_-)E^b\\
            &\hspace{.5cm}-\frac 1{R}\omega^{bc}\bar\theta_-\gamma_{bc}\gamma^{a0}\theta_+\,,\\[.1truecm]
    {\omega}^{ab}&=\frac1{2 x^2}(x^b dx^a- x^adx^b)\Big(\cosh \frac xR-1 \Big)\,,\\[.1truecm]
    {E_-}_\alpha&=[d\theta_-]_\alpha
                   -\frac1{2R}[\gamma_a\theta_+]_\alpha E^a + \frac12\omega^{ab}[\gamma_{ab}\theta_-]_\alpha\,,\\[.1truecm]
    {E_+}_\alpha&=[d\theta_+]_\alpha
                  -\frac1{2R}[\gamma_a\theta_-]_\alpha E^a + \frac12\omega^{ab}[\gamma_{ab}\theta_+]_\alpha \,,\\[.1truecm]
    E_Z &= ds
           + \frac12 \bar \theta_-\gamma^0d\theta_-
           - \frac12 \bar \theta_+\gamma^0d\theta_+\\[.1truecm]
          &\hspace{.5cm} -\frac 1{2}\omega^{ab}(\bar\theta_+\gamma_{ab}\gamma^0\theta_+
                          +\bar\theta_-\gamma_{ab}\gamma^0\theta_-)
                          +\frac 1{2R}\bar\theta_-\gamma^{a0}\theta_+
                           E^a\,.
\end{aligned}
\end{equation}
We can use the supervielbein to write the Maurer-Cartan form of the $\mathcal N=(2,0)$ Carroll superparticle as follows:
\begin{equation}
\begin{aligned}
\label{3DLsN2AdS}
    L_H& = E^0
           + \frac12 v_aE^a \,,\hspace{2.3cm}
    L_P^a =E^a \,,\\
    L^a_K &= \omega^{a0}+dv^a+2v_b\,\omega^{ab}\,,\hspace{1.15cm}
    L_Z = E_Z\,,\\
    {L_-}_\alpha&={E_-}_\alpha\,,\hspace{3.45cm}
    {L_+}_\alpha={E_+}_\alpha\,.
\end{aligned}
\end{equation}

\subsection{Global Symmetries and Kappa symmetry}

The action of the Carrollian superparticle is given by the pull-back of all $L$'s
that are invariant under rotations:
\begin{equation}
\label{AdSAction3D}
\begin{aligned}
    S&=a\int(L_H)^*+b\int(L_Z)^* \\[.1truecm]
     &=a\int d\tau \left( \dot t \cosh\frac xR+\frac R{2x}v_a\dot x^a\sinh\frac xR
         +\frac1{2x^2}x^bv_b x_a \dot x^a \Big(1-\frac Rx\sinh\frac xR\Big) \right.\\[.1truecm]
         &\hspace{2.2cm}
         - \frac14 \bar\theta_-\gamma^0 \dot\theta_-
         - \frac14 \bar\theta_+\gamma^0 \dot\theta_+
         -\frac 1{8x^2}x^b\dot x^a (\bar\theta_+\gamma_{ab}\gamma^0\theta_+
         +\bar\theta_-\gamma_{ab}\gamma^0\theta_-)\Big(\cosh\frac xR -1\Big)\\[.1truecm]
         &\hspace{2.2cm}\left.+\frac 1{4x}\bar\theta_-\gamma^{a0}\theta_+
         \Big[\dot x_a\sinh \frac xR + \frac 1{Rx}x_ax_b\dot x^b
         \Big(1-\frac Rx\sinh\frac xR\Big)\Big] \right) \\[.1truecm]
     &\hspace{.25cm}+b\int d\tau \left( \dot s
            + \frac12 \bar \theta_-\gamma^0\dot\theta_-
            - \frac12 \bar \theta_+\gamma^0\dot\theta_+\right.
            -\frac 1{4x^2}x^b\dot x^a(\bar\theta_+\gamma_{ab}\gamma^0\theta_+
            -\bar\theta_-\gamma_{ab}\gamma^0\theta_-)\Big(\cosh\frac xR -1\Big)\\[.1truecm]
            &\hspace{2.2cm}\left.+\frac 1{2x}\bar\theta_-\gamma^{a0}\theta_+
            \Big[\dot x_a\sinh \frac xR + \frac 1{Rx}x_ax_b\dot x^b
            \Big(1-\frac Rx\sinh\frac xR\Big)\Big] \right) \,,
\end{aligned}
\end{equation}
which is invariant under the following bosonic transformation rules for the coordinates with constant parameters
$(\zeta$, $\eta$,   $a^i$, $\lambda^i$, $\lambda^i_j)$
corresponding to time translations, $Z$ transformations, spatial translations, boosts, rotations, respectively
\begin{equation}
\begin{aligned}
    \delta t&=-\zeta+\frac R{2x} \lambda^k x_k \tanh \frac xR
              +\frac t{Rx}a^k x_k\tanh\frac xR\,,\\[.1truecm]
    \delta x^i&= -\frac 1{x^2}\left(x^ia^k x_k
                 -\frac xR \coth\frac xR (x^ia^kx_k-a^ix^2)\right)
                 -2\lambda^{i}_{\,k}\,x^k \,,\\[.1truecm]
    \delta s &=-\eta\,,\\
    \delta v^i&= -\lambda^i-\frac1{x^2}\lambda^kx_kx^i\text{sech}\frac xR
     \left(1-\cosh \frac xR\right)
               -2\lambda^{i}_{\,j}\,v^j\\[.1truecm]
               &\hspace{.4cm}-\frac{2t}{R^2}a^i-\frac{2t}{R^2x^2}x^ia^kx_k\,\text{sech}\frac xR\left(1-\cosh\frac xR\right)
               +\frac2{Rx}v_ba^{[i}x^{b]}\text{csch}\frac xR \left( 1-\cosh\frac xR\right)
               \,,\\[.1truecm]
    \delta \theta_+ &= -\frac 12 \lambda^{ab}\gamma_{ab}\theta_+
                      +\frac1{2Rx}a^kx^b\gamma_{kb}\theta_- \text{csch}\frac xR
                      \left( 1-\cosh\frac xR\right)\,,\\[.1truecm]
    \delta \theta_- &= -\frac 12 \lambda^{ab}\gamma_{ab}\theta_-
                       +\frac1{2Rx}a^kx^b\gamma_{kb}\theta_+ \text{csch}\frac xR
                        \left( 1-\cosh\frac xR\right)\,.
\end{aligned}
\end{equation}
The same action is invariant under fermionic transformation rules with constant parameters $(\epsilon_+, \epsilon_-)$ corresponding to the supersymmetry transformations
\begin{equation}
\begin{aligned}
    \delta t&=\frac14\bar\epsilon_+\gamma^0\theta_+ \text{sech} \frac xR \cosh \frac x{2R}
              -\frac1{4x}x^k\bar\epsilon_+\gamma^{k0}\theta_-
                   \text{sech} \frac xR \sinh \frac x{2R}\\[.1truecm]
             &\hspace{.4cm}
             +\frac14\bar\epsilon_-\gamma^0\theta_- \text{sech} \frac xR \cosh \frac x{2R}
             -\frac1{4x}x^k\bar\epsilon_-\gamma^{k0}\theta_+
                   \text{sech} \frac xR \sinh \frac x{2R}\,,\\[.1truecm]
    \delta x^i&= 0 \,,\\
    \delta v^i&= \frac 1{Rx} x^i \bar \epsilon_+ \gamma^0\theta_+
                \Big(\frac12 \tanh\frac xR\cosh \frac x{2R}-2\sinh\frac x{2R}\Big)
                +\frac 1R \bar \epsilon_+\gamma^{i0}\theta_-\cosh \frac x{2R}\\[.1truecm]
                &\hspace{.4cm}-\frac 1{2Rx^2}x^ix^k\bar \epsilon_+\gamma^{k0}\theta_-
                \tanh\frac xR\sinh \frac x{2R}\\[.1truecm]
                &\hspace{.4cm}
                +\frac 1{2Rx} x^i \bar \epsilon_- \gamma^0\theta_-
                \tanh\frac xR\cosh \frac x{2R}
                -\frac 1{Rx}x^b \bar \epsilon_- \gamma_b\gamma^{i0}\theta_-\sinh\frac x{2R}
               \\[.1truecm]
                &\hspace{.4cm}-\frac 1{2Rx^2}x^ix^k\bar \epsilon_-\gamma^{k0}\theta_+
                                \tanh\frac xR\sinh \frac x{2R}\,,\\[.1truecm]
    \delta s &=\frac12\bar\epsilon_+\gamma^0\theta_+ \cosh \frac x{2R}
                  +\frac1{2x}x^k\bar\epsilon_+\gamma^{k0}\theta_-  \sinh \frac x{2R}\\
                 &\hspace{.4cm}
                 -\frac12\bar\epsilon_-\gamma^0\theta_-  \cosh \frac x{2R}
                 -\frac1{2x}x^k\bar\epsilon_-\gamma^{k0}\theta_+  \sinh \frac x{2R} \,,\\[.1truecm]
    \delta \theta_+ &= \epsilon_+\cosh\frac x{2R}
    +\frac1x x^k\gamma_k\epsilon_-\sinh\frac x{2R}\,,\\[.1truecm]
    \delta \theta_- &=\epsilon_-\cosh\frac x{2R}
                     +\frac 1x x^k\gamma_k\epsilon_+\sinh\frac x{2R}\,.\\[.1truecm]
\end{aligned}
\end{equation}

To derive an action that is invariant under $\kappa$-transformations we need to
find a fermionic gauge-transformation that leaves $L_H$ and/or $L_Z$ invariant.
The variation of $L_H$ and $L_Z$ under gauge-transformations are given by
\begin{equation}
\label{gaugesym}
\begin{aligned}
   \delta L_H&=d([\delta z_H])+\frac12L_P^a[\delta z_K^a]+\frac12L_K^a[\delta z_P^a]
               +\frac12\bar L_-\gamma^0[\delta z_-]
               +\frac12 \bar L_+\gamma^0[\delta z_+]\,,\\[.1truecm]
   \delta L_Z&= d([\delta z_Z])-\bar L_-\gamma^0[\delta z_-]+\bar L_+\gamma^0[\delta z_+]\,,
\end{aligned}
\end{equation}
where, for example, $[\delta z_K^a]$ is obtained from $L_H$ by changing the 1-forms $dt,$ $ d\theta_+,$  $d\theta_-$ with the transformations $\delta t,$ $\delta\theta_+,$  $\delta\theta_-$. For $\kappa$-transformations we have $[\delta z_H]=0,$ $[\delta z_K^a]=0,$ $[\delta z_P^a]=0$ and hence
 we find
\begin{equation}
\begin{aligned}
    \delta L_H&=\frac12\delta\bar \theta_-\gamma^0[\delta z_-]
                    +\frac12\delta\bar \theta_+\gamma^0[\delta z_+] \,,\\[.1truecm]
    \delta L_Z&=-\delta\bar \theta_-\gamma^0[\delta z_-]
                    +\delta\bar \theta_+\gamma^0[\delta z_+]\,.
\end{aligned}
\end{equation}
It follows that to obtain a $\kappa$-symmetric action we need to take the pull-back
of  either $L_H$ or $L_Z$, with $b=\pm\frac12a$. We focus here on the case $b=-\frac12 a$.
For this choice the action and $\kappa$-symmetry rules are  given by
\begin{equation}
\begin{aligned}
    S&=a\int (L_H-\frac12 L_Z)^*\,, \qquad [\delta z_+]=\kappa\,,\qquad [\delta z_-]=0\,,
\end{aligned}
\end{equation}
\noindent where $\kappa=\kappa(\tau)$ is an arbitrary local parameter.
Using this we find
the following $\kappa$-transformations of the coordinates
\begin{equation}
\begin{aligned}
    \delta t&=\frac14\text{sech}\frac xR\bar \theta_+\gamma^0\kappa\,,\hspace{.65cm}
    \delta x^a=0\,,\hspace{4.15cm}
    \delta\theta_+=\kappa\,,\\[.1truecm]
    \delta s&=\frac12\bar \theta_+\gamma^0\kappa\,,\hspace{1.8cm}
    \delta v_a =\frac 1{2Rx} x^a\bar \theta_+ \gamma^0\kappa \tanh\frac xR\,,\hspace{.8cm}
    \delta \theta_-=0\,.
\end{aligned}
\end{equation}
After $\kappa$-gauge fixing (setting $\theta_+=0$) the action reads
\begin{equation}
\label{AdSAction3Dkfix}
\begin{aligned}
    S&=a\int d\tau \left( \dot t \cosh\frac xR-\frac12\dot s
         +\frac R{2x}v_a\dot x^a\sinh\frac xR
         +\frac1{2x^2}x^bv_b x_a \dot x^a \Big(1-\frac Rx\sinh\frac xR\Big) \right.\\
         &\hspace{2.2cm}
         \left.- \frac12 \bar\theta_-\gamma^0 \dot\theta_-
         -\frac 1{4x^2}x^b\dot x^a
         \bar\theta_-\gamma_{ab}\gamma^0\theta_-\Big(\cosh\frac xR -1\Big)\right)\,.
\end{aligned}
\end{equation}
This action is invariant under the following transformation rules
\begin{equation}
\begin{aligned}
\label{TransfN2Kfix}
    \delta t&=-\frac1{4x}x^k\bar\epsilon_+\gamma^{k0}\theta_-
              \text{sech} \frac xR \sinh \frac x{2R}
             +\frac14\bar\epsilon_-\gamma^0\theta_-\text{sech}\frac xR\cosh\frac x{2R}\,,\\
    \delta x^i&= 0 \,,\\
    \delta v^i&=\frac 1R \bar \epsilon_+\gamma^{i0}\theta_-\cosh \frac x{2R}
                -\frac 1{2Rx^2}x^ix^k\bar \epsilon_+\gamma^{k0}\theta_-
                \tanh\frac xR\sinh \frac x{2R}\\
                &\hspace{.4cm}
                +\frac 1{2Rx} x^i \bar \epsilon_- \gamma^0\theta_-
                \tanh\frac xR\cosh \frac x{2R}
                -\frac 1{Rx}x^b \bar \epsilon_- \gamma_b\gamma^{i0}\theta_-\sinh\frac x{2R}
               \\
    \delta s &=-\frac1{2x}x^k\bar\epsilon_+\gamma^{k0}\theta_-  \sinh \frac x{2R}
                 -\frac12\bar\epsilon_-\gamma^0\theta_-\cosh\frac x{2R}\,,\\
    \delta \theta_- &=\epsilon_-\cosh\frac x{2R}
                     +\frac 1x x^k\gamma_k\epsilon_+\sinh\frac x{2R}\,.\\
\end{aligned}
\end{equation}

\subsection{The \texorpdfstring{$\mathcal N=(1,1)$}{N=(1,1)} AdS Carroll Superalgebra}

We now consider the $3D$ $\mathcal{N}=(1,1)$ anti-de Sitter algebra which is given by
\begin{equation}
\begin{aligned}
    \left[M_{AB}, M_{CD}\right]&=2\eta_{A[C}M_{D]B}-2\eta_{B[C}M_{D]A}\,,\hspace{2.45cm}
    [M_{AB},Q^\pm]=-\frac12\gamma_{AB}Q^\pm\,,\\[.1truecm]
    [M_{AB},P_C]&=2\eta_{C[B}P_{A]}\,,\hspace{5.6cm}
    [P_A,Q^\pm]=\pm x\gamma_A Q^\pm\,,\\[.1truecm]
    \{Q^\pm_\alpha,Q^\pm_\beta\}&=4[\gamma^AC^{-1}]_{\alpha\beta}P_A
                              \pm 4x[\gamma^{AB}C^{-1}]_{\alpha\beta}M_{AB}\,,\hspace{1cm}
   [P_A,P_B]=4x^2M_{AB}\,.
\end{aligned}
\end{equation}
Here $P_A,M_{AB}$ and $Q^\pm_\alpha$ are the generators of space-time translations, Lorentz rotations and supersymmetry transformations, respectively. The bosonic generators $P_A$ and $M_{AB}$ are anti-hermitian while de fermionic generators $Q^\pm_\alpha$ are hermitian. Like in the previous case, the parameter $x=1/(2R)$ is a contraction parameter.

To make the Carroll contraction we rescale the generators with a parameter $\omega$ as follows:
\begin{equation}
\begin{aligned}
    P_0&=\frac\omega2 H\,,\hspace{1cm}
    M_{a0}=\omega K_a\,,\hspace{1cm}
    Q^\pm=\sqrt\omega\tilde Q^\pm\,.
\end{aligned}
\end{equation}
Taking the limit $\omega\rightarrow\infty$ and dropping the tildes on the $Q^\pm$ we get the following 3D $\mathcal N=(1,1)$ Carroll superalgebra:
\begin{equation}
\begin{aligned}
    \label{AdSCarrAlgN11}
    \left[M_{ab},P_c\right]&=2\delta_{c[b}P_{a]}\,, \hspace{1cm}
    [M_{ab},K_c]=2\delta_{c[b}K_{a]}\,, \\[.1truecm]
    [P_a, P_b]&=\frac1{R^2}M_{ab}\,,\hspace{1cm}
    [P_a,K_b]=\frac12\delta_{ab}H\,,\hspace{1cm}
    [P_a,H]=\frac 2{R^2}K_a\\[.1truecm]
    [P_a,Q^\pm]&=\pm\frac 1{2R}\gamma_a Q^\pm\,,\hspace{1cm}
    [M_{ab},Q^\pm]=-\frac 12 \gamma_{ab}Q^\pm\,,\\[.1truecm]
    \{ Q^\pm_\alpha, Q^\pm_\beta\}&=2[\gamma^0C^{-1}]H
                                   \pm \frac 4R [\gamma^{a0}C^{-1}]_{\alpha\beta}K_a\,.
\end{aligned}
\end{equation}
The corresponing componetns of the Maurer-Cartan equation $dL^C-\frac12\,{f^C}_{AB}L^BL^A=0$ are given by
\begin{equation}
\begin{aligned}
    dL_H&= -\frac12 L_P^{\;a}\,L_K^{\;a} - \bar L_+\gamma^0L_+ - \bar L_-\gamma^0L_- \,,\\
    d L^{\;a}_P&= 2L_P^{\;b}\,L_M^{\;ab}\,,\\[.1truecm]
    d L^{\;a}_K&= 2L_K^{\;b}\,L_M^{\;ab}+\frac2{R^2}L_H{L_P}^a
                  -\frac2R\bar L_+\gamma^{a0}L_+
                  +\frac2R\bar L_-\gamma^{a0}L_-\,, \\[.1truecm]
    d L_M^{\;ab}&=2L_M^{\;ca}\,L_M^{\;cb}+\frac1{2R^2}{L_P}^b{L_P}^a\,,\\[.1truecm]
    d L_+&=\frac12 \gamma_{ab} L_+ \,L_M^{\;ab}-\frac1{2R}\gamma_aL_+{L_P}^a\,,\\[.1truecm]
    d L_-&=\frac12 \gamma_{ab} L_- \,L_M^{\;ab}+\frac1{2R}\gamma_aL_-{L_P}^a\,.
\end{aligned}
\end{equation}

\subsection{Superparticle Action}

Taking the algebra  \eqref{AdSCarrAlgN11}  we consider the following coset
\begin{equation}
 \frac G H=\frac {\mathcal N=(1,1)\textrm { AdS Carroll}}{\textrm {SO(D-1)}}\,.
\end{equation}
The coset element is $g= g_0\,U,$ where $g_0=e^{Ht} e^{ P_a x^a} e^{Q_\alpha^-\theta_-^\alpha} e^{Q_\alpha^+\theta_+^\alpha}$ is the coset representing  the $\mathcal N=(1,1)$ Carroll superspace and $U=e^{ K_a v^a}$ is a general Carroll boost representing the insertion of the superparticle..

The Maurer-Cartan form associated to the super-Carroll space is given by
\begin{equation}
\Omega_0=(g_0)^{-1}dg_0=H E^0+P_aE^a + K_a \omega^{a0}+
M_{ab}\omega^{ab} - \bar Q^- E_- - \bar Q^+ E_+ \,,
\end{equation}
where $(E^0, E^a, {E_-}_{\alpha},{E_+}_{\alpha})$ and $(\omega^{a0},\omega^{ab})$ are the supervielbein and the spin connection of the Carroll superspace:
\begin{equation}
\begin{aligned}
    E^0& =  d t \cosh\frac xR
           -  \bar\theta_-\gamma^0 d\theta_-
           -  \bar\theta_+\gamma^0 d\theta_+
           - \omega^{ab} (\bar\theta_+\gamma_{ab}\gamma^0\theta_+
                                      +\bar\theta_-\gamma_{ab}\gamma^0\theta_-) \,,\\[.1truecm]
    E^a &=\frac Rxdx^a\sinh\frac xR
                  +\frac 1{x^2}x^ax^bdx_b\Big(1-\frac Rx \sinh \frac xR\Big) \,,\\[.1truecm]
    {\omega}^{a0} &= -\frac 2{xR}dtx^a\sinh\frac xR
                    -\frac1{R^2}(\bar\theta_+\gamma_{ab}\gamma^0\theta_+
            +\bar\theta_-\gamma_{ab}\gamma^0\theta_-)E^b\\[.1truecm]
           &\hspace{.5cm}  -\frac 2R(\bar\theta_+\gamma^{a0}d\theta_+ - \bar \theta_- \gamma^{a0} d\theta_- )\,,\\[.1truecm]
    {\omega}^{ab}&=\frac1{2 x^2}(x^b dx^a- x^adx^b)\Big(\cosh \frac xR-1 \Big)\,,\\
    {E_-}_\alpha&=[d\theta_-]_\alpha
                   +\frac1{2R}[\gamma_a\theta_-]_\alpha E^a + \frac12\omega^{ab}[\gamma_{ab}\theta_-]_\alpha\,,\\[.1truecm]
    {E_+}_\alpha&=[d\theta_+]_\alpha
                  -\frac1{2R}[\gamma_a\theta_+]_\alpha E^a + \frac12\omega^{ab}[\gamma_{ab}\theta_+]_\alpha \,.\\[.1truecm]
\end{aligned}
\end{equation}
We can use the supervielbein to write the Maurer-Cartan form of the $\mathcal N=(1,1)$ Carroll superparticle as follows:
\begin{equation}
\begin{aligned}
    L_H& = E^0
           + \frac12 v_aE^a \,,\hspace{2.3cm}
    L_P^a =E^a \,,\\
    L^a_K &= \omega^{a0}+dv^a+2v_b\,\omega^{ab}\,,\\
    {L_-}_\alpha&={E_-}_\alpha\,,\hspace{3.45cm}
    {L_+}_\alpha={E_+}_\alpha\,.
\end{aligned}
\end{equation}

\subsection{Global Symmetries}

The action of the Carrollian superparticle is given by the pull-back of all $L$'s
that are invariant under rotations:
\begin{equation}
\label{AdSAction3DN11}
\begin{aligned}
    S&=M\int(L_H)^* \\
     &=M\int d\tau \left( \dot t \cosh\frac xR+\frac R{2x}v_a\dot x^a\sinh\frac xR
         +\frac1{2x^2}x^bv_b x_a \dot x^a \Big(1-\frac Rx\sinh\frac xR\Big) \right.\\[.1truecm]
         &\hspace{.2cm}\left.
         -  \bar\theta_-\gamma^0 \dot\theta_-
         -  \bar\theta_+\gamma^0 \dot\theta_+
         -\frac 1{2x^2}x^b\dot x^a (\bar\theta_+\gamma_{ab}\gamma^0\theta_+
         +\bar\theta_-\gamma_{ab}\gamma^0\theta_-)\Big(\cosh\frac xR -1\Big) \right) \,.\\
\end{aligned}
\end{equation}
This action is invariant under the following bosonic transformation rules for the coordinates with constant parameters
$(\zeta$,   $a^i$, $\lambda^i$, $\lambda^i_j)$
corresponding to time translations, spatial translations, boosts and rotations, respectively
\begin{equation}
\begin{aligned}
\label{AdSBosTransfN11}
    \delta t&=-\zeta+\frac R{2x} \lambda^k x_k \tanh \frac xR
              +\frac t{Rx}a^k x_k\tanh\frac xR\,,\\[.1truecm]
    \delta x^i&= -\frac 1{x^2}\left(x^ia^k x_k
                 -\frac xR \coth\frac xR (x^ia^kx_k-a^ix^2)\right)
                 -2\lambda^{i}_{\,k}\,x^k \,,\\[.1truecm]
    \delta v^i&= -\lambda^i-\frac1{x^2}\lambda^kx_kx^i\text{sech}\frac xR
     \left(1-\cosh \frac xR\right)
               -2\lambda^{i}_{\,j}\,v^j\\[.1truecm]
               &\hspace{.4cm}-\frac{2t}{R^2}a^i-\frac{2t}{R^2x^2}x^ia^kx_k\,
                  \text{sech}\frac xR\left(1-\cosh\frac xR\right)\\[.1truecm]
              &\hspace{.5cm} +\frac2{Rx}v_ba^{[i}x^{b]}\text{csch}\frac xR \left( 1-\cosh\frac xR\right)
               \,,\\[.1truecm]
    \delta \theta_+ &= -\frac 12 \lambda^{ab}\gamma_{ab}\theta_+
                      +\frac1{2Rx}a^kx^b\gamma_{kb}\theta_+ \text{csch}\frac xR
                      \left( 1-\cosh\frac xR\right)\,,\\[.1truecm]
    \delta \theta_- &= -\frac 12 \lambda^{ab}\gamma_{ab}\theta_-
                         +\frac1{2Rx}a^kx^b\gamma_{kb}\theta_- \text{csch}\frac xR
                         \left( 1-\cosh\frac xR\right)\,.
\end{aligned}
\end{equation}
The same action is invariant inder the following fermionic transformation rules with constant parameters $(\epsilon_+, \epsilon_-)$ corresponding to  supersymmetry transformations:
\begin{equation}
\begin{aligned}
    \delta t&=\bar\epsilon_+\gamma^0\theta_+ \text{sech} \frac xR \cosh \frac x{2R}
              -\frac1{x}x^k\bar\epsilon_+\gamma^{k0}\theta_+
                   \text{sech} \frac xR \sinh \frac x{2R}\\[.1truecm]
             &\hspace{.4cm}
             +\bar\epsilon_-\gamma^0\theta_- \text{sech} \frac xR \cosh \frac x{2R}
             +\frac1{x}x^k\bar\epsilon_-\gamma^{k0}\theta_-
                   \text{sech} \frac xR \sinh \frac x{2R}\,,\\[.1truecm]
    \delta x^i&= 0 \,,\\[.1truecm]
    \delta v^i&=\frac 2R\bar \epsilon_+\gamma^{i0}\theta_+\cosh\frac x{2R}
                 -\frac 2{Rx}x^k\bar \epsilon_+\gamma_k\gamma^{i0}\theta_+\sinh \frac x{2R}\\[.1truecm]
                &\hspace{.4cm}  +\frac 2{xR}x^i \tanh\frac xR
                   \Big(\bar\epsilon_+\gamma^0\theta_+\cosh\frac x{2R}
                   -\frac1x x^k\bar \epsilon_+ \gamma^{k0}\theta_+\sinh\frac x{2R}\Big)\\[.1truecm]
                &\hspace{.4cm} -\frac 2R\bar \epsilon_-\gamma^{i0}\theta_-\cosh\frac x{2R}
                 -\frac 2{Rx}x^k\bar \epsilon_-\gamma_k\gamma^{i0}\theta_-\sinh \frac x{2R}\\[.1truecm]
                &\hspace{.4cm}  +\frac 2{xR}x^i \tanh\frac xR
                    \Big(\bar\epsilon_-\gamma^0\theta_-\cosh\frac x{2R}
                     +\frac1x x^k\bar \epsilon_- \gamma^{k0}\theta_-\sinh\frac x{2R}\Big)\\[.1truecm]
    \delta \theta_+ &= \epsilon_+\cosh\frac x{2R}
    +\frac1x x^k\gamma_k\epsilon_+\sinh\frac x{2R}\,,\\[.1truecm]
    \delta \theta_- &=\epsilon_-\cosh\frac x{2R}
                     -\frac 1x x^k\gamma_k\epsilon_-\sinh\frac x{2R}\,.
\end{aligned}
\end{equation}

\bibliography{CarrollSuperparticle}
\bibliographystyle{utphys}

\end{document}